%% file: main.tex
\documentclass[aps,prd,twocolumn,superscriptaddress,tightenlines,nofootinbib]{revtex4-1}
\usepackage[T1]{fontenc}
\usepackage{microtype}
\usepackage[textsize=scriptsize,backgroundcolor=red!70,linecolor=red]{todonotes}
\usepackage{booktabs}
\usepackage{amsmath}
\usepackage{amsfonts}
\usepackage{amssymb}
\usepackage{hyperref}
\usepackage[all]{hypcap}
\usepackage{paralist}
\usepackage{multirow}
\usepackage{graphicx}
\usepackage{dcolumn}
\usepackage{bm}
\usepackage{epsf}

\newcommand*{\rom}[1]{\expandafter\@slowromancap\romannumeral #1@}

\def\be{\begin{equation}}
\def\ee{\end{equation}}
\def\bea{\begin{eqnarray}}
\def\eea{\end{eqnarray}}
\def\gsim{\ \rlap{\raise 2pt\hbox{$>$}}{\lower 2pt \hbox{$\sim$}}\ }
\def\lsim{\ \rlap{\raise 2pt\hbox{$<$}}{\lower 2pt \hbox{$\sim$}}\ }
\def\dslash{\kern-4pt \not{\hbox{\kern-2pt $\partial$}}}
\def\pslash{\not{\hbox{\kern-2pt p}}}

\def\beq{\begin{equation}}
\def\eeq{\end{equation}}

\def\bctaunu{b \to c \tau^- {\bar\nu}}

\def\bcellnuX{b \to c \ell^- {\bar X}}

\def \cB{{\cal B}}

\def \be{\beta}

\def \({\left(}
\def \){\right)}
\def \[{\left[}
\def \]{\right]}

\def\beq{\begin{equation}}
\def\eeq{\end{equation}}
\def\beqa{\begin{eqnarray}}
\def\eeqa{\end{eqnarray}}

\def\ra{\rightarrow}

\def\beq{\begin{equation}}
\def\eeq{\end{equation}}
\def\beqa{\begin{eqnarray}}
\def\eeqa{\end{eqnarray}}

\begin{document}
	\DeclareGraphicsExtensions{.eps,.ps}
	
	
	\title{$\bar B \to  D^{(*)} \ell \bar{X}$  decays in effective field theory with massive right-handed neutrinos}
	
	\author{Alakabha Datta}
	\email[Email Address: ]{datta@phy.olemiss.edu}
	\affiliation{Department of Physics and Astronomy, University of Mississippi, Oxford, MS 38677, USA}
	
	\author{Hongkai Liu}
	\email[Email Address: ]{liu.hongkai@campus.technion.ac.il}
	\affiliation{Department of Physics, Technion – Israel Institute of Technology, Haifa 3200003, Israel}
	
	\author{Danny Marfatia}
	\email[Email Address: ]{dmarf8@hawaii.edu}
	\affiliation{Department of Physics and Astronomy, University of Hawaii at Manoa, Honolulu, HI 96822, USA}
	\affiliation{Kavli Institute for Theoretical Physics, University of California, Santa Barbara, CA 93106, USA}
	
	\begin{abstract}
		
		We calculate the complete differential decay distributions for the $B$ meson decays, $\bar B \to  D^{(*)} \ell \bar{X}$, to a massive right-handed (RH) neutrino in the low-energy effective field theory (LEFT) framework.
		We find that a massive RH neutrino does not introduce any new angular structures compared to the massless case, but can cause significant distortions in angular observables.  
		 We  study the phenomenology of low-energy four-fermion operators permitted by the standard model effective field theory (SMEFT) extended with RH neutrinos (SMNEFT).  We show that to explain the positive value of the difference in forward-backward asymmetries, $\Delta A_{\text{FB}}\equiv A_{\text{FB}}^\mu-A_{\text{FB}}^e$, tentatively inferred from Belle data, the RH neutrino must be massive. We also make predictions for $q^2$ dependent  angular observables to motivate future measurements. 
		
	\end{abstract}
	
\maketitle

{\bf Introduction.} 
Hints of new physics (NP) have been reported in the charged current decays, $B \to D^{(*)} \tau \nu_\tau$,  by the BaBar, Belle and LHCb experiments. Measurements of the ratios, $R_{D^{(*)}}^{\tau/\ell} \equiv \cB(\bar{B} \to D^{(*)} \tau^{-} {\bar\nu}_\tau)/\cB(\bar{B} \to D^{(*)} \ell^{-}{\bar\nu}_\ell)$, where $\ell = e,\mu$, are larger than the standard model (SM) predictions~\cite{BaBar:2012obs, BaBar:2013mob, LHCb:2015gmp, Belle:2015qfa, Belle:2016ure, Belle:2016dyj, LHCb:2017smo, Belle:2017ilt, LHCb:2017rln, Belle:2019gij} with a combined  significance of $3.4\sigma$ \cite{HFLAV:2019otj}.  This is known as the
$R_{D^{(*)}}$ puzzle. Measurements of a similar ratio, $R_{J/\psi}^{\tau/\mu} \equiv \cB(B_c^+ \to J/\psi\tau^+\nu_\tau) / \cB(B_c^+ \to J/\psi\mu^+\nu_\mu)$ \cite{LHCb:2017vlu}, also show tension with the SM at 1.7$\sigma$ significance  \cite{Watanabe:2017mip}. 
These measurements suggest NP in $\bctaunu$ decays that is lepton universality violating (LUV). 

Not surprisingly, most of 
 the theoretical work on NP has been concentrated on semileptonic $\tau$ modes  with a  left-handed (LH) neutrino in the final state.
 If NP allows for decays to a  light right-handed (RH) neutrino, the decay rate is always enhanced because there is no interference with the SM amplitude in the limit of vanishing active neutrino mass. This feature can be
 used to naturally explain the $R_{D^*}^{\tau/ \ell}$ measurements~\cite{He:2012zp, Cvetic:2017gkt, Asadi:2018wea, Greljo:2018ogz, Babu:2018vrl, Mandal:2020htr}.  However, 
 with the limited experimental statistics, it is difficult to find clear signals of NP. Moreover,
since the final state contains one or more additional neutrinos from $\tau$ decay, measurements of angular distributions that are crucial for detecting NP are further complicated. 
   

In the coming years, the $B$ factories, Belle II and LHCb,
may conclusively confirm the existence of beyond the standard model (SM)  physics in semileptonic $B$ decays. In this Letter, we study the high statistics charged current
$\bar B \to  D^{(*)} \ell \bar{X}$ decay arising from the underlying  $\bcellnuX$ transition, where $ \ell = e, \mu$ and the invisible state $X$ can be a LH neutrino or a light RH singlet neutrino. 
At Belle~II with 50 ab$^{-1}$ we expect $8\times 10^6$ events in each of the muon and electron modes.
These modes allow full event reconstruction because the missing neutrino momentum can be calculated from the $e^+e^-$ kinematics at the $\Upsilon(4S)$.

New physics in the muon sector is motivated by anomalies in the measured value of 
$(g-2)_\mu$~\cite{Muong-2:2021ojo} and neutral-current $b \to s \mu^+\mu^-$ decays~\cite{LHCb:2021trn}. 
Since our interest is in LUV NP, we assume NP to affect only the muon sector while the electron sector is described by the SM. In this spirit we introduce a RH neutrino associated with the muon.
LUV NP in the electron and muon sectors is tightly constrained by the measurement of the ratio of rates, $R_{D^{(*)}}^{\mu/e} \equiv \cB(\bar{B} \to D^{(*)} \mu^{-} {\bar\nu}_\mu)/\cB(\bar{B} \to D^{(*)} e^{-} {\bar\nu}_e)$ which is within 5\% of unity. We restrict ourselves to NP scenarios in which this ratio can deviate up to 3\% from unity, a precision achievable in the future. 

A key point is that even if the effects of LUV NP are small in the ratios of decay rates, larger effects may be visible in the angular distributions as functions of $q^2$, and
angular observables  may provide one or more unambiguous signals for NP. One of the issues that should be addressed is whether form factor uncertainties can obscure these signals. Fortunately, we can identify two types of observables in the SM  that have very little or no form factor uncertainties and hence any measured deviations from the SM predictions for these observables would be clear signs of NP. The first are the
 $\Delta$ observables that quantify differences in the angular observables for the muon and electron channels, e.g.,  $\Delta A_{FB}  \equiv A_{FB}^\mu  - A_{FB}^e $, where $A_{FB}$  is  the forward-backward asymmetry. The second type are the  CP-violating triple-product terms in the angular distribution~\cite{Duraisamy:2013pia,Duraisamy:2014sna} which can be nonzero if
  NP couplings are complex and have phases different from the SM contribution. The measurements of CP violating terms require large statistics~\cite{Bhattacharya:2022cna}, and so we focus on the $\Delta$ observables in this work.
Recently, using the tables of Belle data in Ref.~\cite{Belle:2018ezy}, an anomaly in $\Delta A_{FB}$ was reported in Ref.~\cite{Bobeth:2021lya}. If confirmed, this could signal LUV~\cite{Bobeth:2021lya,Carvunis:2021dss,Bhattacharya:2022cna}. As an application of our formalism, we explore if decays to a massive RH neutrino can resolve this anomaly.

Note that while  the effects of a right-handed neutrino have been considered in the
 $\tau$ channel, our approach has several novel features. (1) In our framework, the structure of the low energy effective operators is assumed to  arise from the
 standard model effective field theory (SMEFT) extended with RH neutrinos (SMNEFT). Consequently, only the subset of operators compatible with this well motivated formalism for physics above the electroweak scale, is allowed. (2) We present, for the first time, the complete angular distribution
 for $\bar B \to  D^{(*)} \ell \bar{X}$  decays with a massive right-handed neutrino. (3) We address the 
 $\langle\Delta A_{FB} \rangle$ anomaly  with the aid of a massive RH neutrino, and show that a massless RH neutrino fails to do so.



{\bf SMNEFT.} 
Standard model effective field theory (SMEFT)~\cite{Grzadkowski:2010es,Henning:2014wua, Brivio:2017vri} 
is defined in terms of $SU(3)_C \times SU(2)_L \times U(1)_Y$ invariant higher dimensional operators ${\cal O}^i$ built from SM fields:
\begin{eqnarray}
{\cal L}  =  \sum_{i} \frac{c_i}{\Lambda^{d_i-4}} {\cal O}^i , \ 
\end{eqnarray}
where $\Lambda$ is the NP scale above the electroweak scale, $d_i>4$ are integer dimensions of ${\cal O}^i$, and the dimensionless
parameters $c_i$ are the Wilson's coefficients (WCs) that
can be calculated by matching the effective theory with the underlying
theory. 

\begin{table}
	\centering
	\begin{tabular}{| c   c   c   c || c  c  c c | }
		\toprule
		$\mathcal{O}^{(3)}_{\ell q}$  & $\mathcal{O}^{(1)}_{\ell equ}$& $\mathcal{O}_{\ell edq}$  & $\mathcal{O}^{(3)}_{\ell eqd}$ & $\mathcal{O}_{nedu}$  & $\mathcal{O}_{\ell nuq}$& $\mathcal{O}^{(1)}_{\ell nqd}$  & $\mathcal{O}^{(3)}_{\ell nqd}$   \\
		\midrule
		$\mathcal{O}^V_{LL}$  & $\mathcal{O}^S_{LL}$& $\mathcal{O}^S_{RL}$  & $\mathcal{O}^T_{LL}$ & $\mathcal{O}^V_{RR}$  & $\mathcal{O}^S_{LR}$& $\mathcal{O}^S_{RR}$  & $\mathcal{O}^T_{RR}$    \\
		\midrule
	\end{tabular}
	\caption{The origin of low-energy effective operators from SMNEFT. The last four operators in the second row arise by extending SMEFT to SMNEFT.}
	\label{Table: op}
\end{table}

Motivated by neutrino mass and oscillations, RH neutrinos that 
 that are sterile under the SM gauge interactions can be incorporated into SMEFT.  The resulting EFT~\cite{delAguila:2008ir, Aparici:2009fh, Bhattacharya:2015vja, Liao:2016qyd, Bischer:2019ttk}, called SMNEFT, includes additional interactions of the RH neutrinos with SM fields. The mass scale of the RH neutrino can vary over a large range. We consider the case of a light RH neutrino so that it appears as an explicit degree of freedom in the EFT framework.

{\bf \boldmath{$\bar B \to  D^{(*)} \ell \bar{X}$}.} In a general EFT at the $m_b$ scale,  NP in semileptonic $B$ decays can be described  by four-fermion contact interactions that give $b\ra c\ell\bar{X}$. The dimension-six $SU(3)_C \times U(1)_Q$ invariant Lagrangian is
\beq
-{\cal L}_{\text{eff}} = \frac{4G_FV_{cb}}{\sqrt{2}}(\mathcal{O}^V_{LL}+\sum_{\substack{X=S,V,T\\\alpha,\beta=L,R}}C^X_{\alpha\beta}~\mathcal{O}^X_{\alpha\beta})\,,
\label{eq:wc}
\eeq
where
\beqa
\mathcal{O}^V_{\alpha\beta} &\equiv& (\bar{c}\gamma^\mu P_\alpha b) (\bar{\ell}\gamma^\mu P_\beta \nu)\,,\\
\mathcal{O}^S_{\alpha\beta} &\equiv& (\bar{c} P_\alpha b) (\bar{\ell} P_\beta \nu)\,,\\
\mathcal{O}^T_{\alpha\beta} &\equiv& \delta_{\alpha\beta} (\bar{c} \sigma^{\mu\nu} P_\alpha b) (\bar{\ell}\sigma_{\mu\nu} P_\beta \nu)\,.
\eeqa
The first term in Eq.~(\ref{eq:wc}) is the SM contribution, and the NP is in the second term.
As these operators should emerge from SMNEFT, the two EFTs must match at the electroweak scale. From SMEFT, only the operators $\mathcal{O}^V_{LL},\,\mathcal{O}^S_{LL},\,\mathcal{O}^S_{RL}$, and $\mathcal{O}^T_{LL}$ arise, while SMNEFT yields four more operators: $\mathcal{O}^V_{RR},\,\mathcal{O}^S_{LR},\,\mathcal{O}^S_{RR}$, and $\mathcal{O}^T_{RR}$; see Table~\ref{Table: op}. Note that $\mathcal{O}^V_{LR}$ and $\mathcal{O}^V_{RL}$ cannot be produced from the four-fermion operators in SMNEFT. 	
The renormalization group running of the operators from $\Lambda$ to $m_Z$ and then down to the $m_b$ scale has been discussed in Refs.~\cite{Datta:2020ocb,Datta:2021akg}.  The scalar operator $\mathcal{O}^{(1)}_{\ell nqd}$ and the  tensor operator $\mathcal{O}^{(3)}_{\ell nqd}$ mix via the weak gauge couplings above the weak scale.  Below the weak scale the operators  $\mathcal{O}^S_{RR}$ and $\mathcal{O}^T_{RR}$ mix due to the electromagnetic interaction. The operators on the left and right side of the partition in Table~\ref{Table: op}  mix via Yukawa couplings. In what follows, we work in the low-energy effective field theory (LEFT) framework keeping in mind that the corresponding SMNEFT WCs can be obtained by carrying out the running and matching. 

\begin{figure}[t]
 	\includegraphics[width=0.45\textwidth]{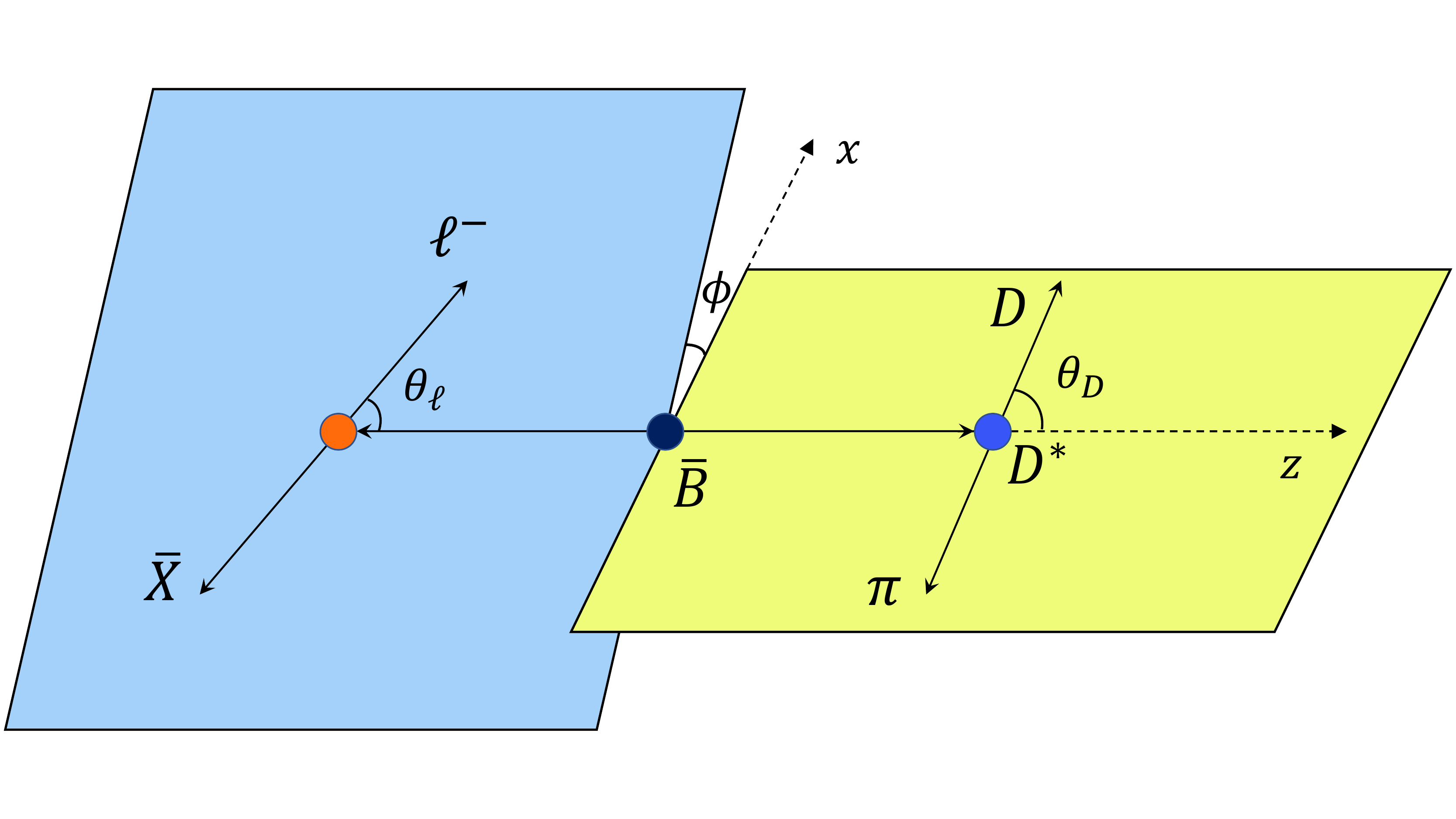}
 	\caption{Kinematic variables for $\bar B\ra \ell^-\bar N D^*(\ra D\pi)$.}
 	\label{fig:decay}
 \end{figure}

{\bf Formalism.} The differential decay distribution for $\bar B\ra D \ell \bar X$ with a massless RH neutrino is given in Ref.~\cite{Mandal:2020htr}. We generalize the result for a nonzero RH neutrino mass $m_N$. A finite $m_N$ affects both the phase space and the leptonic helicity amplitudes. 
For example, the operators with $\beta =R$ produce left-handed antineutrinos with helicity $\lambda_{\bar N} = \pm 1/2$ because the mass flips the helicity. 
The differential decay distribution  for $\bar B\ra D \ell \bar X$  can be expressed in terms of the three $\mathcal{J}$ functions as    
\beqa
\frac{d^2\Gamma_D}{dq^2d\cos\theta_{\ell}} &=&  \sum_{i=0}^2 \mathcal{J}_i(q^2,\vec{\bf{C}}, m_N) \tilde{f}_i(\cos\theta_{\ell})\,,
\eeqa
where $q^2 \equiv (p_{\ell}+p_{\bar N})^2$ and  $\tilde{f}_i(\cos\theta_{\ell})$ are the angular functions with $\theta_{\ell}$ the angle between the charged lepton momentum in the  $\ell\bar X$ rest frame
and the direction of the $D$ momentum in the $\bar B$ rest frame.
 The $\mathcal{J}_i$ functions depend on $q^2$, WCs $\vec{\bf{C}}$ and  $m_N$, and
are provided in {\it Supplemental Material}. 
Similarly, the differential decay distribution for $\bar B\ra D^* (\ra D\pi)\ell \bar X$ with nonzero $m_N$, can be written in terms of the 12 different angular structures that appear in the massless RH neutrino case:
\beqa
\frac{d^4\Gamma_{D^*}}{dq^2d\cos\theta_{\ell}d\cos\theta_Dd\phi} =\frac{3}{8\pi} \sum_i\mathcal{I}_i\,f_i(\cos\theta_{\ell},\cos\theta_D,\phi),\nonumber\\
\label{Dstardist}
\eeqa
where $\mathcal{I}_i\equiv \mathcal{I}_i(q^2,\vec{\bf{C}}, m_N)$, and the three angles are defined in Fig.~\ref{fig:decay}; our convention for $\theta_\ell$ differs from that often used by experimentalists~\cite{Bhattacharya:2022cna}. For the complete expression see {\it Supplemental Material}.
For $m_N = 0$, our $\mathcal{I}$ and $\mathcal{J}$ functions match the $I$ and $J$ functions of Ref.~\cite{Mandal:2020htr}.
We adopt the hadronic form factors of Ref.~\cite{Bordone:2019guc} including the corrections up to $1/m_c^2$ in the heavy-quark limit. 

{\bf Phenomenology.} The general angular distributions can be integrated over  subsets of the variables to construct several distributions. The differential distributions with respect to $q^2$ are
\beqa
\Gamma^D_f(q^2)\equiv\frac{d\Gamma_{D}}{dq^2} &=& 2 \mathcal{J}_0(q^2) + \frac{2}{3} \mathcal{J}_2(q^2)\,,\\
\Gamma^{D^*}_f(q^2)\equiv\frac{d\Gamma_{D^*}}{dq^2} &=& 2\mathcal{I}_{1s}(q^2) +\mathcal{I}_{1c}(q^2) \nonumber\\
&&-\frac{1}{3} (2 \mathcal{I}_{2s}(q^2)+\mathcal{I}_{2c}(q^2))\,.\nonumber\\
\eeqa
We define 9 bins of the normalized $q^2$-distributions~\cite{Bobeth:2021lya},
\beqa
\Delta x^{D^{(*)}}_{i} \equiv \frac{1}{\Gamma_{\text{tot}}^{D^{(*)}}}\int_{q^2_{i-1}}^{q^2_{i}}dq^2 \Gamma^{D^{(*)}}_f(q^2),\quad i = 2~\text{to}~10\,,
\eeqa
where $\Gamma^{D^{(*)}}_{\text{tot}}$ is the total decay width after integrating $\Gamma^{D^{(*)}}_f(q^2)$ over the entire range of $q^2$. The $q^2$ bins are defined by
\beq
q^2_i \equiv m_B^2 + m_{D^{(*)}}^2-2 m_ B m_{D^{(*)}} \omega_i\,,\quad i= 1~\text{to}~10\,,
\eeq 
with $\omega_i = 1 + i/20$.
 The differential distributions with respect to $\cos\theta_{\ell}$, $\cos\theta_{D}$, and $\phi$ after integrating over the other variables, can be written in terms of five angular observables $\langle A^{D^*}_{\text{FB}}\rangle$, 
 $\langle \tilde{F}_L\rangle$, $\langle F_L\rangle$, $\langle S_3\rangle$, and $\langle S_9\rangle$:
\beqa
\frac{1}{\Gamma^{D^*}_{\text{tot}}}\frac{d\Gamma^{D^*}}{d\cos\theta_{\ell}} &= &\frac{1}{2} - \langle A^{D^*}_{\text{FB}}\rangle \cos\theta_{\ell}\nonumber\\
&&+\frac{1}{4}(1-3\langle \tilde{F}_L\rangle)\frac{3\cos^2\theta_{\ell}-1}{2}\,,\\
\frac{1}{\Gamma^{D^*}_{\text{tot}}}\frac{d\Gamma^{D^*}}{d\cos\theta_{D}} &=& \frac{3}{4}[1-\langle F_L\rangle +(3 \langle F_L\rangle - 1)\cos^2\theta_{D}],\\
\frac{1}{\Gamma^{D^*}_{\text{tot}}}\frac{d\Gamma^{D^*}}{d\phi} &=&\frac{1}{2\pi} +\frac{2}{3\pi}\langle S_3\rangle\cos(2\phi) \nonumber\\
&&+ \frac{2}{3\pi}\langle S_9\rangle\sin(2\phi)\,,
\eeqa
where the $q^2$-averaged observables are defined by
\beq
\langle O\rangle\equiv \frac{1}{\Gamma^{D^{(*)}}_{\text{tot}}}\int_{q^2_{\text{min}}}^{q^2_{\text{max}}}dq^2 O(q^2)\Gamma^{D^{(*)}}_f(q^2)\,.
\eeq
The values of $\langle A^{D^*}_{\text{FB}}\rangle, \langle \tilde{F}_L\rangle, \langle F_L\rangle$ and $\langle S_3\rangle$, measured by the Belle experiment are listed in Table~\ref{tab:obs_meas}. 
Measurements of the two ratios of branching fractions $R^{\mu/e}_{D^{(*)}}$
are also listed in Table~\ref{tab:obs_meas}.
Several additional $q^2$ dependent angular asymmetries can be extracted from the full angular distribution  through asymmetric integrals:
\begin{widetext}
	\bea
	S_4(q^2) &=& \frac{3\pi}{8}(\int_{0}^{1}-\int_{-1}^{0})d\cos\theta_{\ell}(\int^{1}_{0}-\int^{0}_{-1})d\cos\theta_{D}(\int^{\frac{\pi}{2}}_{0}-\int^{\pi}_{\frac{\pi}{2}}-\int^{\frac{3\pi}{2}}_{\pi}-\int_{\frac{3\pi}{2}}^{2\pi})d\phi \frac{d^4\Gamma_{D^*}}{dq^2d\cos\theta_{\ell}d\cos\theta_Dd\phi}\,,\\
	S_5(q^2) &=& \int_{-1}^{1}d\cos\theta_{\ell}(\int^{1}_{0}-\int^{0}_{-1})d\cos\theta_{D}(\int^{\frac{\pi}{2}}_{0}-\int^{\pi}_{\frac{\pi}{2}}-\int^{\frac{3\pi}{2}}_{\pi}-\int_{\frac{3\pi}{2}}^{2\pi})d\phi \frac{d^4\Gamma_{D^*}}{dq^2d\cos\theta_{\ell}d\cos\theta_Dd\phi}\,,\\
	S_7(q^2) &=& \int_{-1}^{1}d\cos\theta_{\ell}(\int^{1}_{0}-\int^{0}_{-1})d\cos\theta_{D}(\int^{\pi}_{0}-\int^{2\pi}_{\pi})d\phi \frac{d^4\Gamma_{D^*}}{dq^2d\cos\theta_{\ell}d\cos\theta_Dd\phi}\,,\\
	S_8(q^2) &=& \frac{3\pi}{8}(\int_{0}^{1}-\int_{-1}^{0})d\cos\theta_{\ell}(\int^{1}_{0}-\int^{0}_{-1})d\cos\theta_{D}(\int^{\pi}_{0}-\int^{2\pi}_{\pi})d\phi \frac{d^4\Gamma_{D^*}}{dq^2d\cos\theta_{\ell}d\cos\theta_Dd\phi}\,.
	\eea
\end{widetext}
In terms of the $\mathcal{I}$ and $\mathcal{J}$ functions, 
\beqa
A^D_{FB} (q^2) &=&  -\frac{\mathcal{J}_1(q^2)}{\Gamma^D_f(q^2)}\,, \\
A^{D^*}_{FB} (q^2) &=& -\frac{\mathcal{I}_{6s}(q^2) + \frac{1}{2} \mathcal{I}_{6c}(q^2)  }{\Gamma^{D^*}_f(q^2)}\,, \\
F_{L}(q^2) &=& \frac{\mathcal{I}_{1c}(q^2) - \frac{1}{3} \mathcal{I}_{2c}(q^2)  }{\Gamma^{D^*}_f(q^2)}\,,\\
\tilde{F}_{L}(q^2) &=& \frac{1}{3} - \frac{8}{9}  \frac{2 \mathcal{I}_{2s}(q^2) + \mathcal{I}_{2c}(q^2)  }{\Gamma^{D^*}_f(q^2)}\,,\\
S_i(q^2) &=& \frac{\mathcal{I}_i(q^2)}{\Gamma^{D^*}_f(q^2)}\,,\quad i =\{3,4,5,7,8,9\}\,.
\eeqa

%
\begin{table}[t]
	\begin{center}
		\begin{tabular}{|c|c|c|c|c|}
			\hline\hline
			Observable  & Measurement & BP1 & BP2 & BP3 \\
			\hline
			$\Delta \langle A^{D^*}_{\text{FB}}\rangle$ & $0.0349 \pm 0.0089$ &0.0188&  -0.0014 & -0.0016 \\
			$\Delta \langle F_{L}\rangle $ & $ -0.0065 \pm 0.0059$ & -0.0057&-0.0063&-0.0025 \\
			$\Delta\langle \tilde{F}_{L}\rangle$ & $-0.0107 \pm 0.0142$&-0.0314&-0.0099&-0.0034 \\
			$\Delta\langle S_{3}\rangle$ & $-0.0127 \pm 0.0109$ & 0.0035&0.0049&0.0007 \\
			$R_{D}^{\mu/e}$ & $0.995 \pm 0.022 \pm 0.039$ &1.015&1.036&1.012 \\
			$R_{D^*}^{\mu/e}$  & $0.99 \pm 0.01 \pm 0.03$ &0.983& 1.021& 0.991 \\
			\hline\hline
			$\Delta x_2^{D^*}$ & $-0.0040 \pm 0.0029$ & -0.0153 & -0.0022 & -0.0002\\
			$\Delta x_3^{D^*}$ & $-0.0025 \pm 0.0033$ & 0.0 & -0.0022 & 0.0001\\
			$\Delta x_4^{D^*}$ & $0.0024 \pm 0.0038$ & 0.0014 & -0.0022 & 0.0002\\
			$\Delta x_5^{D^*}$ & $0.0043\pm 0.0046$ & 0.0022 & -0.0006 & 0.0002\\
			$\Delta x_6^{D^*}$ & $-0.0035 \pm 0.0052$ & 0.0027 & 0.0009 & 0.0003\\
			$\Delta x_7^{D^*}$ & $0.0066 \pm 0.0056$ & 0.0030 & 0.0018 & 0.0003\\
			$\Delta x_8^{D^*}$ & $-0.0103 \pm 0.0054$ & 0.0032 &  0.0021 & 0.0003\\
			$\Delta x_9^{D^*}$ & $0.0 \pm 0.0052$ & 0.0031 & 0.0020 & 0.0003\\
			$\Delta x_{10}^{D^*}$ & $0.0019 \pm 0.0044$ & 0.0028 & 0.0017 & 0.0003\\
			\hline\hline
			$\Delta\langle A^D_{\text{FB}}\rangle$ & - & 0.0401&-0.0032&-0.0209\\
			$\Delta\langle S_4\rangle$ & - & 0.0121&0.0087& 0.0021 \\
			$\Delta\langle S_5\rangle$ & - & -0.0128&-0.0051& 0.0015 \\
			\hline\hline
		\end{tabular}
	\end{center}
	\caption{Ten observables that are sensitive to NP in the $\mu$ sector. The corresponding predictions for the three BPs of Table~\ref{tab:bmp} are provided.}
	\label{tab:obs_meas}
\end{table}

\begin{figure}[t]
	\includegraphics[width=0.47\textwidth]{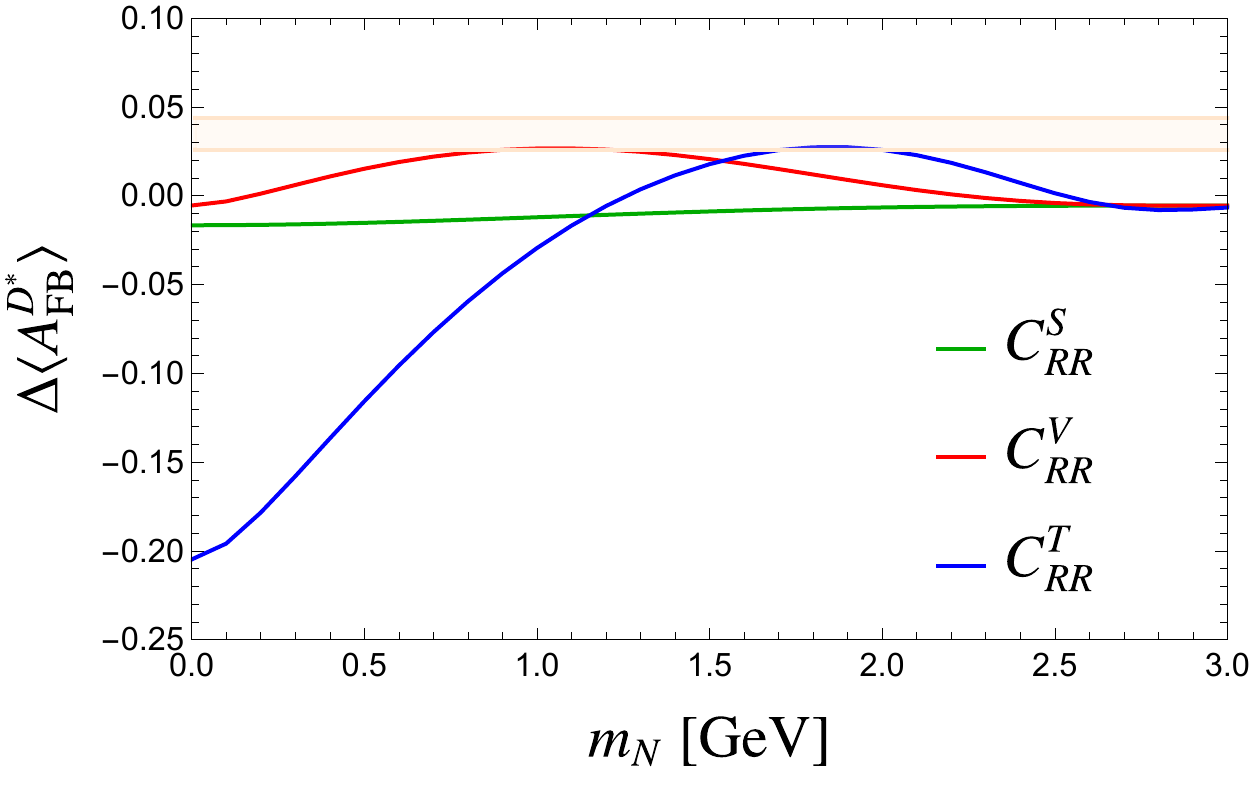}
	\includegraphics[width=0.47\textwidth]{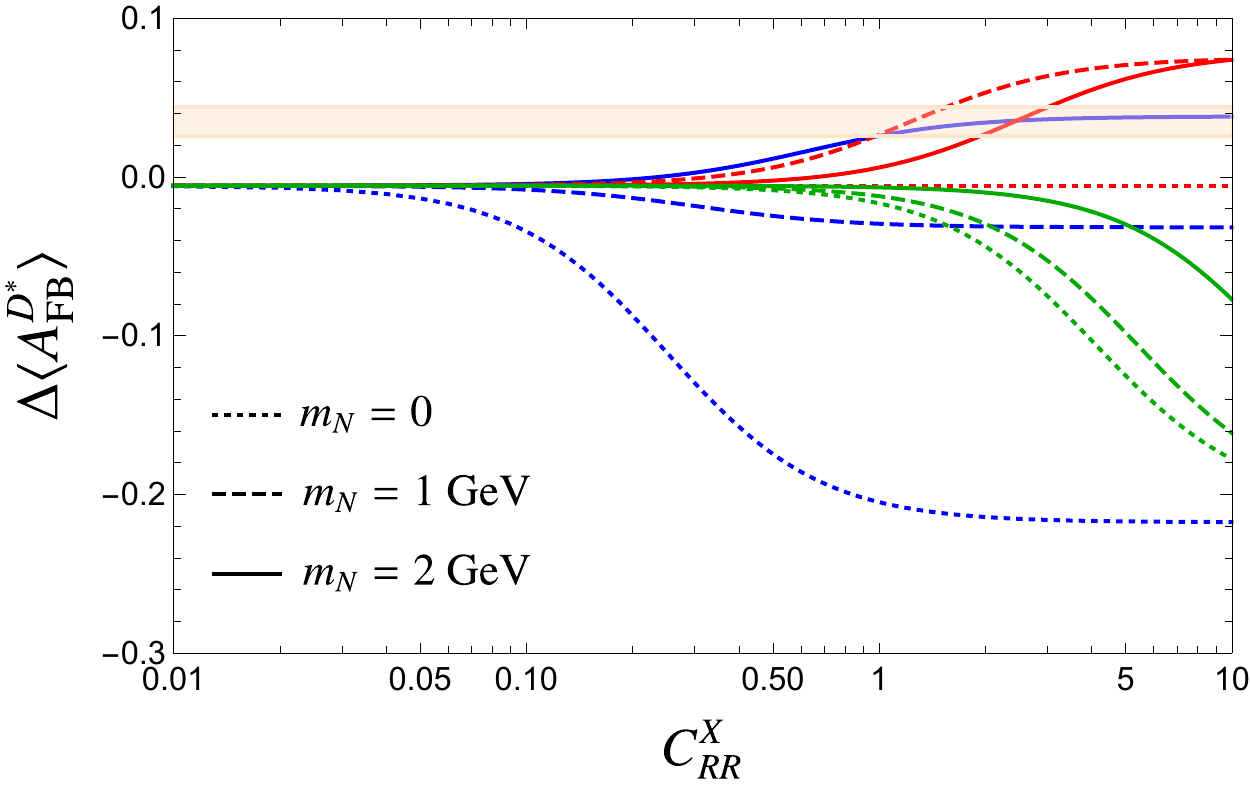}
	\caption{Upper panel: $\Delta \langle A^{D^*}_{\text{FB}}\rangle$ as a function of $m_N$ for $C^S_{RR} = C^V_{RR} = C^T_{RR} = 1$. Lower panel: $\Delta \langle A^{D^*}_{\text{FB}}\rangle$ as a function of $C^S_{RR}$ (green), $C^V_{RR}$ (red), and $C^T_{RR}$ (blue) for $m_N$= 0 (dotted), 1~GeV (dashed), and 2~GeV (solid). The light orange  band shows the Belle measurement at $1\sigma$.}
	\label{fig:afb}
\end{figure}

\begin{table}[h]
	\begin{center}
		\begin{tabular}{c|c|c c c| c c c}
			\hline\hline
			  & $m_N$~(GeV) & $C^V_{RR}$ & $C^S_{RR}$ & $C^T_{RR}$& $C^V_{LL}$& $C^S_{LL}$& $C^T_{LL}$  \\
			\hline
			BP1 & 0.4 & 0.82 & 0.1 & 0.02 & -0.4 & 0 & 0 \\
			BP2   & 1.6 & 0.15 & -0.3 & 0.06 & 0 & 0 & 0   \\
			BP3  & 0 & 0 & 0 & 0 & 0 & 0.06 & 0.02\\
			\hline\hline
		\end{tabular}
	\end{center}
	\caption{The parameters for three benchmark points. The WCs not listed are zero.}
	\label{tab:bmp}
\end{table}

\begin{figure*}[t]
	\centering
	\includegraphics[width=0.3\textwidth]{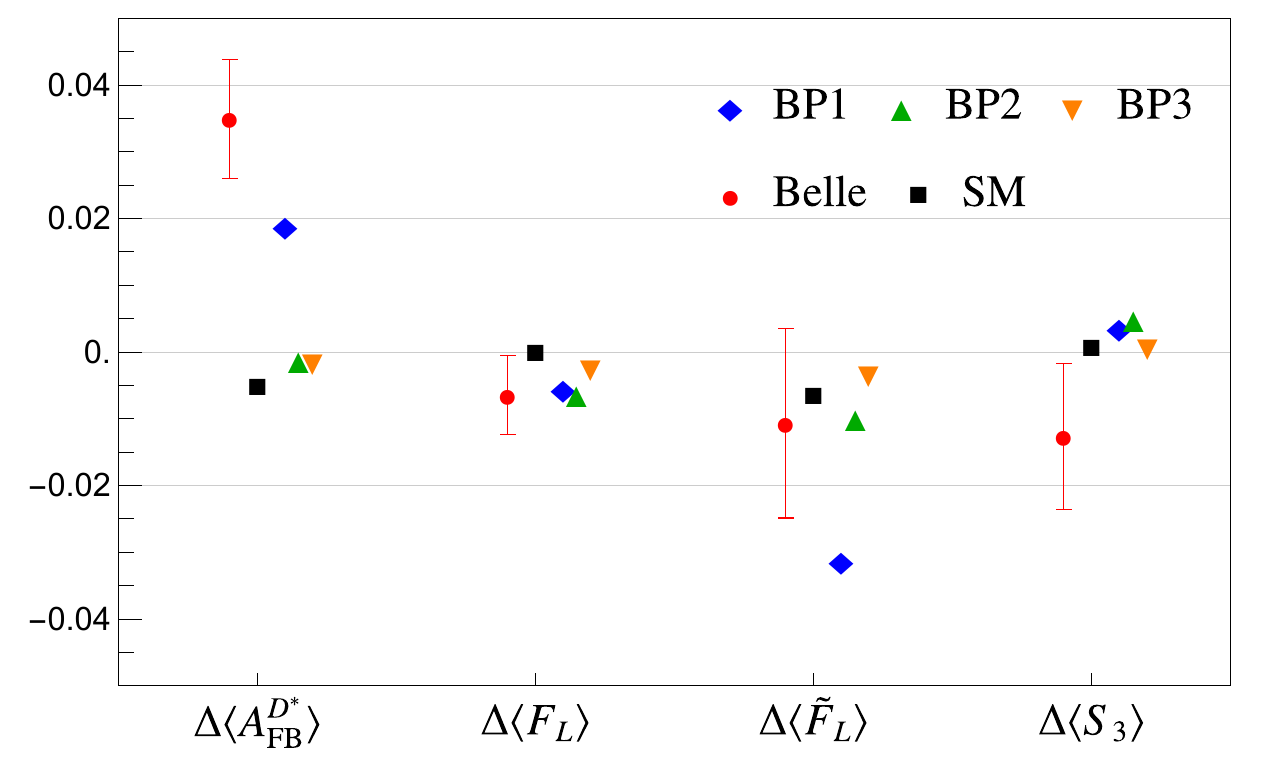}\,
	\includegraphics[width=0.3\textwidth]{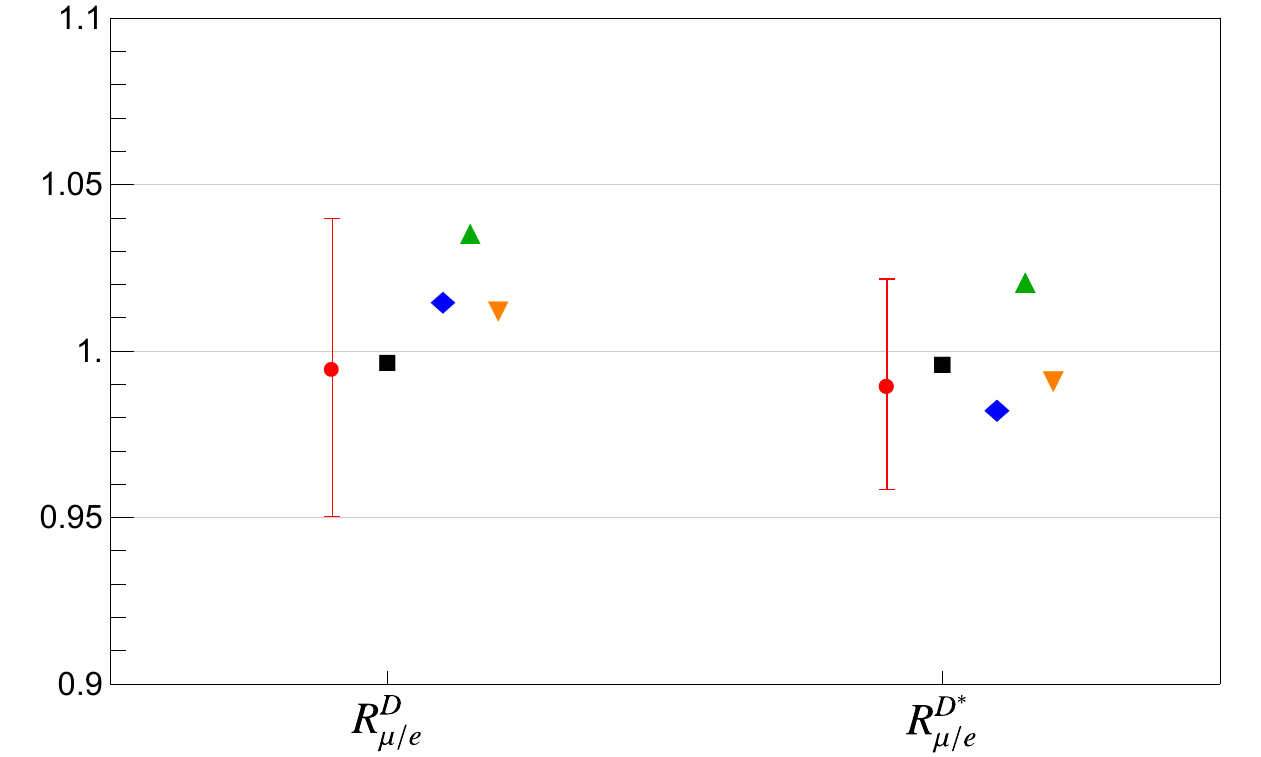}\,
	\includegraphics[width=0.3\textwidth]{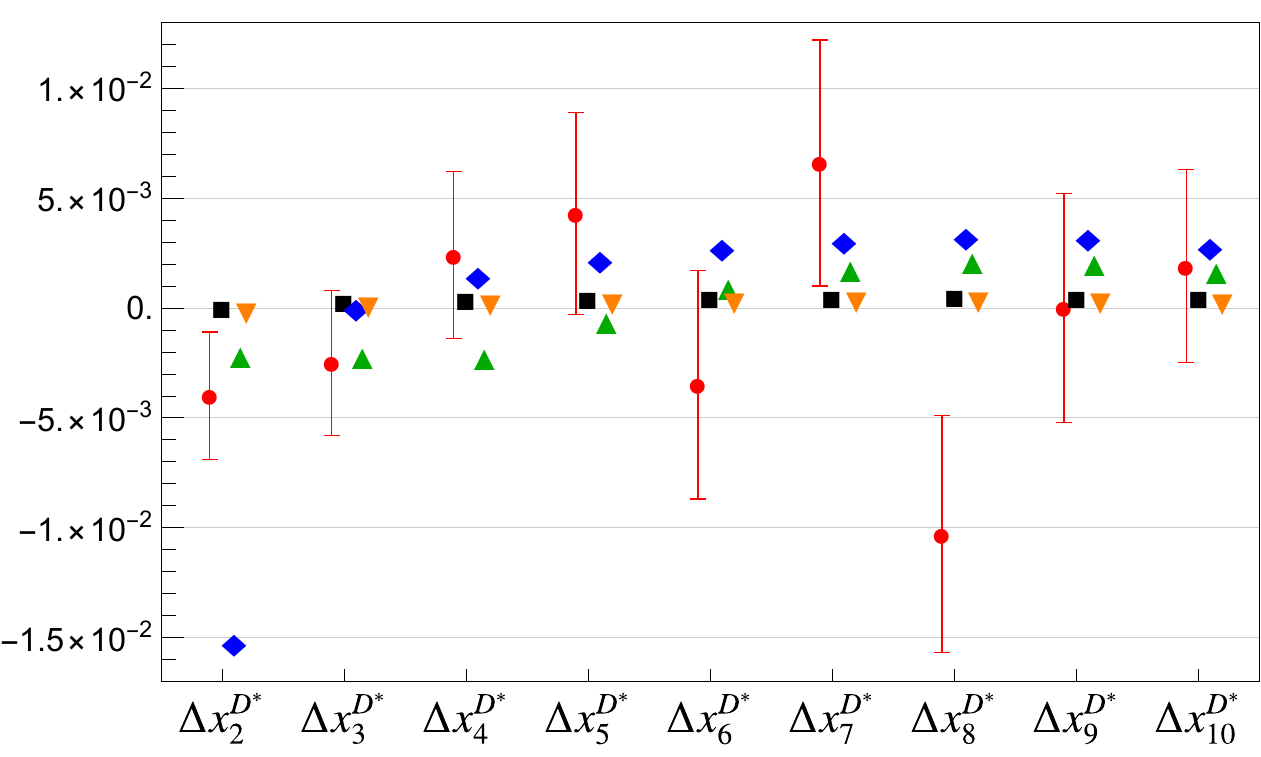}\,
	\includegraphics[width=0.3\textwidth]{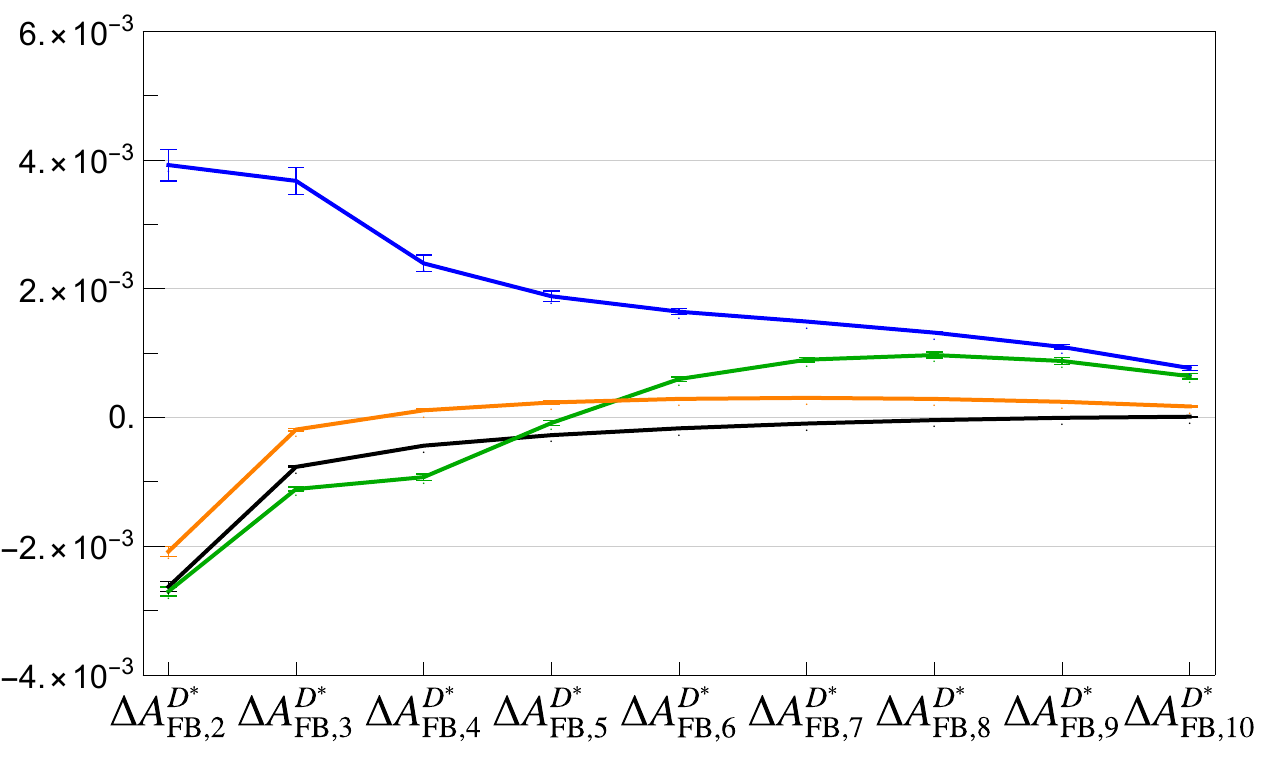}\,
	\includegraphics[width=0.3\textwidth]{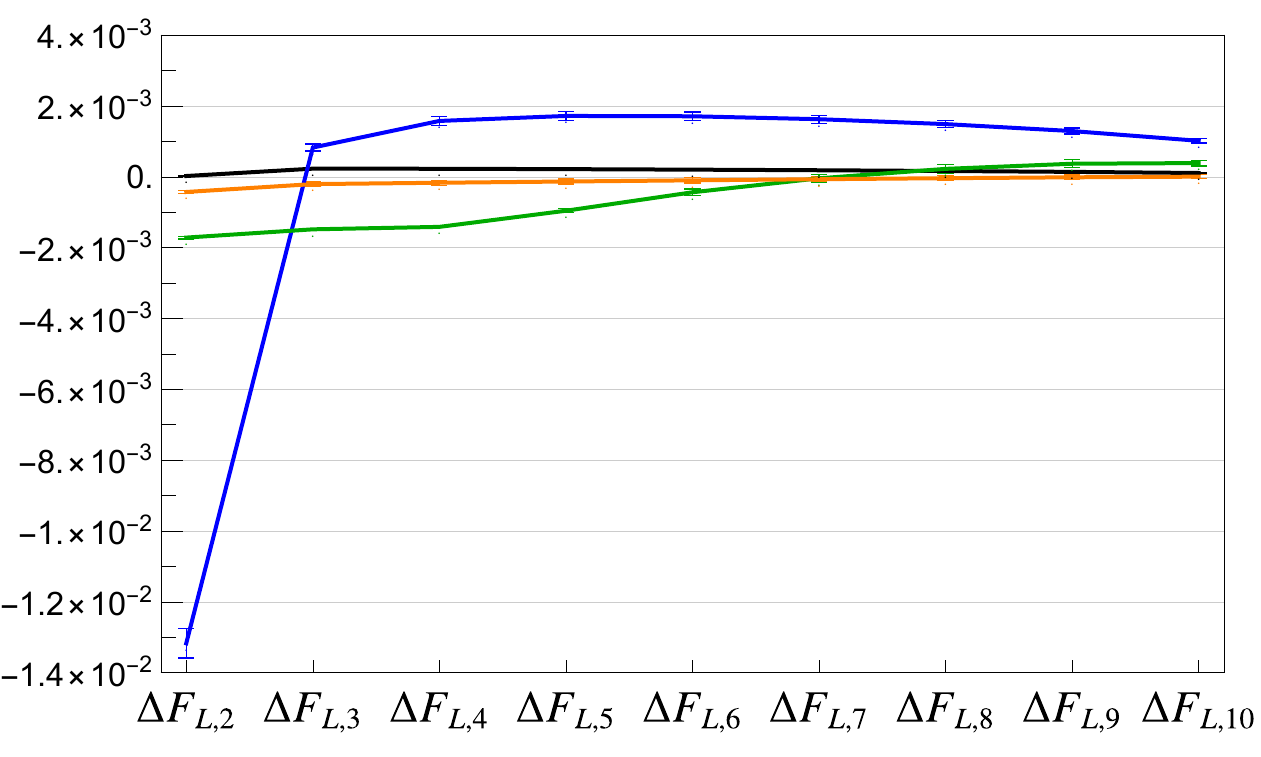}\,
	\includegraphics[width=0.3\textwidth]{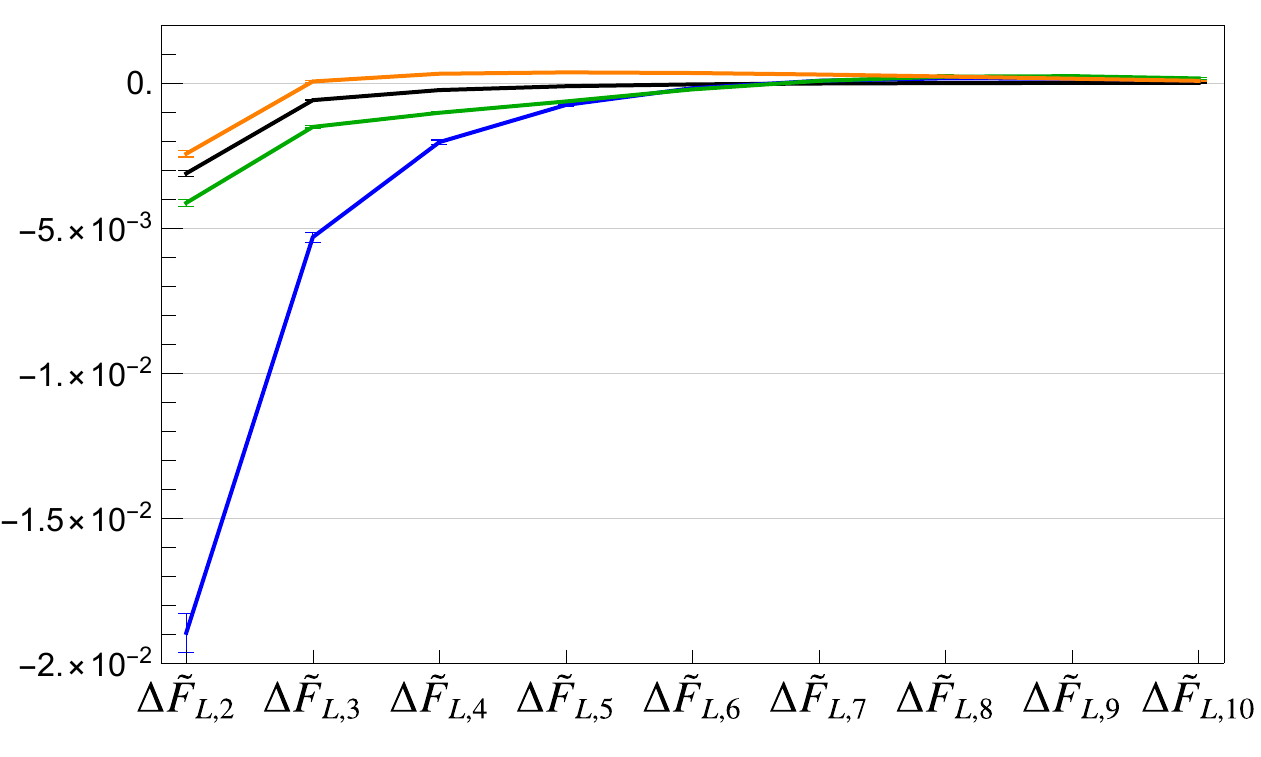}\,
	\includegraphics[width=0.3\textwidth]{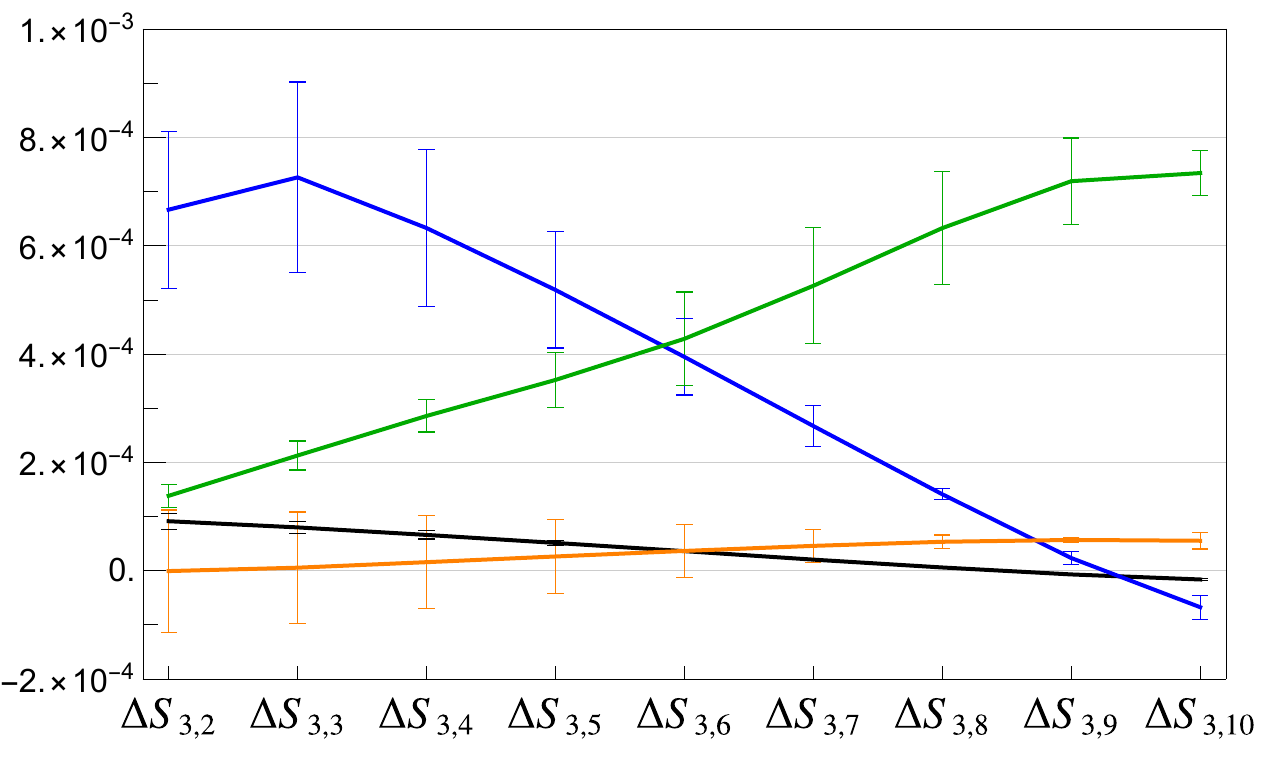}\,
	\includegraphics[width=0.3\textwidth]{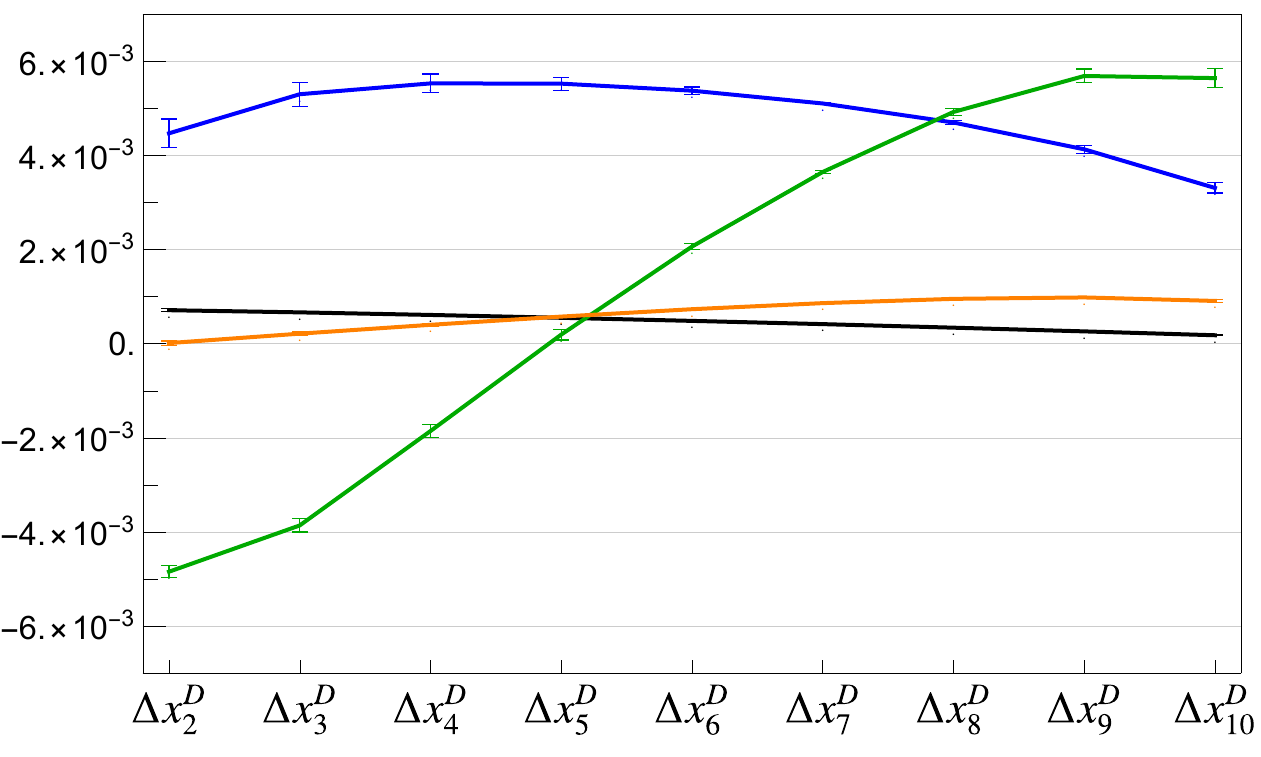}\,
	\caption{The expectations for the SM and three BPs for the observables in the upper and middle panels of Table~\ref{tab:obs_meas}. The Belle measurements are shown in the upper panels. The error bars in the middle and lower panels are hadronic form factor uncertainties. }
	\label{fig:obs}
\end{figure*}

 We find that the nonzero RH neutrino mass produces significant effects in the angular observables which may explain the $4\sigma$ tension in $\Delta \langle A^{D^*}_{\text{FB}}\rangle$. In the upper panel of Fig.~\ref{fig:afb}, we show $\Delta \langle A^{D^*}_{\text{FB}}\rangle$ as a function $m_N$ for $C^S_{RR} = C^V_{RR} = C^T_{RR} = 1$.  Clearly, a GeV RH neutrino with vector or tensor interactions can fit the $\Delta \langle A^{D^*}_{\text{FB}}\rangle$ measurement within 1$\sigma$.
 In the lower panel of Fig.~\ref{fig:afb}, we show the dependence of $\Delta \langle A^{D^*}_{\text{FB}}\rangle$ on the LEFT WCs for three values of $m_N$, taking only one of the WCs to be nonzero at a time. We observe that if the RH neutrino is massless (dotted curves), $\Delta \langle A^{D^*}_{\text{FB}}\rangle$ is always below the SM prediction. However, for $m_N = 1$~GeV, 
 the $\Delta \langle A^{D^*}_{\text{FB}}\rangle$ anomaly can be explained if $C^V_{RR} \approx 1$ (red dashed curve). For $m_N = 2$~GeV, the anomaly can be explained by both $C^V_{RR} \approx 2$ and $C^T_{RR} \approx 1$. However, these illustrative scenarios are excluded by other measurements in Table~\ref{tab:obs_meas}. So, to reproduce the $\Delta \langle A^{D^*}_{\text{FB}}\rangle$ anomaly and the other measurements in Table~\ref{tab:obs_meas}, we choose three benchmark points (BPs) of Table~\ref{tab:bmp}.  BP1 has both LH and RH interactions. while BP2 and BP3 only have RH and LH interactions, respectively. The predictions for the three BPs for the 15 measurements are provided in Table~\ref{tab:obs_meas} and Fig.~\ref{fig:obs}. 
Since there is no interference between LH and RH contributions, scenarios with only RH interactions (like BP2) necessarily increase $R_{D}^{\mu/e}$ and $R_{D^*}^{\mu/e}$, and it is not possible to sufficiently enhance 
$\Delta \langle A^{D^*}_{\text{FB}}\rangle$. Only LH interactions (BP3) are unable to adequately reproduce all the measurements. 
 It is clear that BP1 can alleviate the tension in $\Delta\langle A^{D^*}_{\text{FB}}\rangle$ to within $ \sim 2 \sigma$. This requires a large correction to the vector LH interaction in conjunction with a large contribution from the vector RH neutrino interaction. It is possible to obtain predictions closer to the central values of $\Delta \langle A^{D^*}_{\text{FB}}\rangle$ and $\Delta \tilde{F}_L$, at the expense of an even larger cancellation of $C^V_{LL}$ with the SM. One such set of parameters is $C^V_{LL} = -0.84$, $C^V_{RR} = 1.0$, $C^S_{RR} = 0.05$, $C^T_{RR} = 0.03$, and $m_N = 0.3$~GeV.  
 
\begin{figure*}[t]
	\centering
	\includegraphics[width=0.3\textwidth]{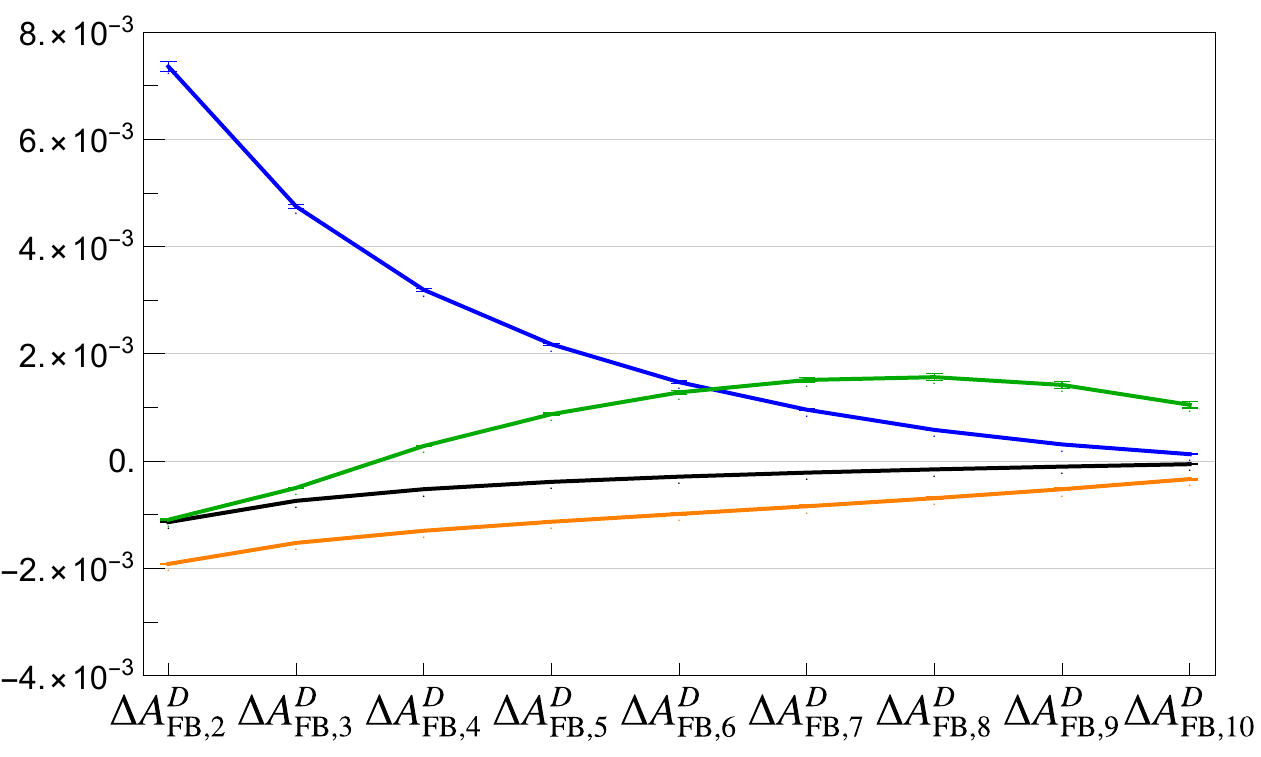}\,
\includegraphics[width=0.3\textwidth]{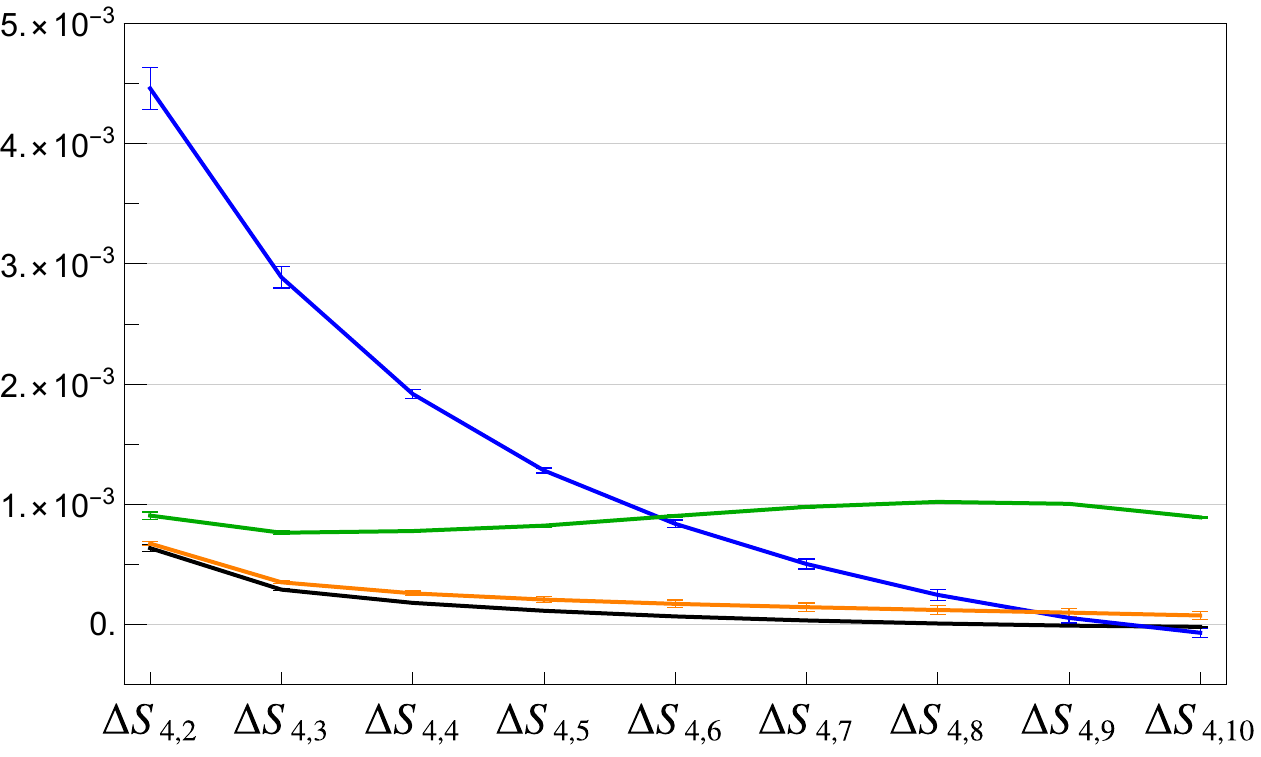}\,
\includegraphics[width=0.3\textwidth]{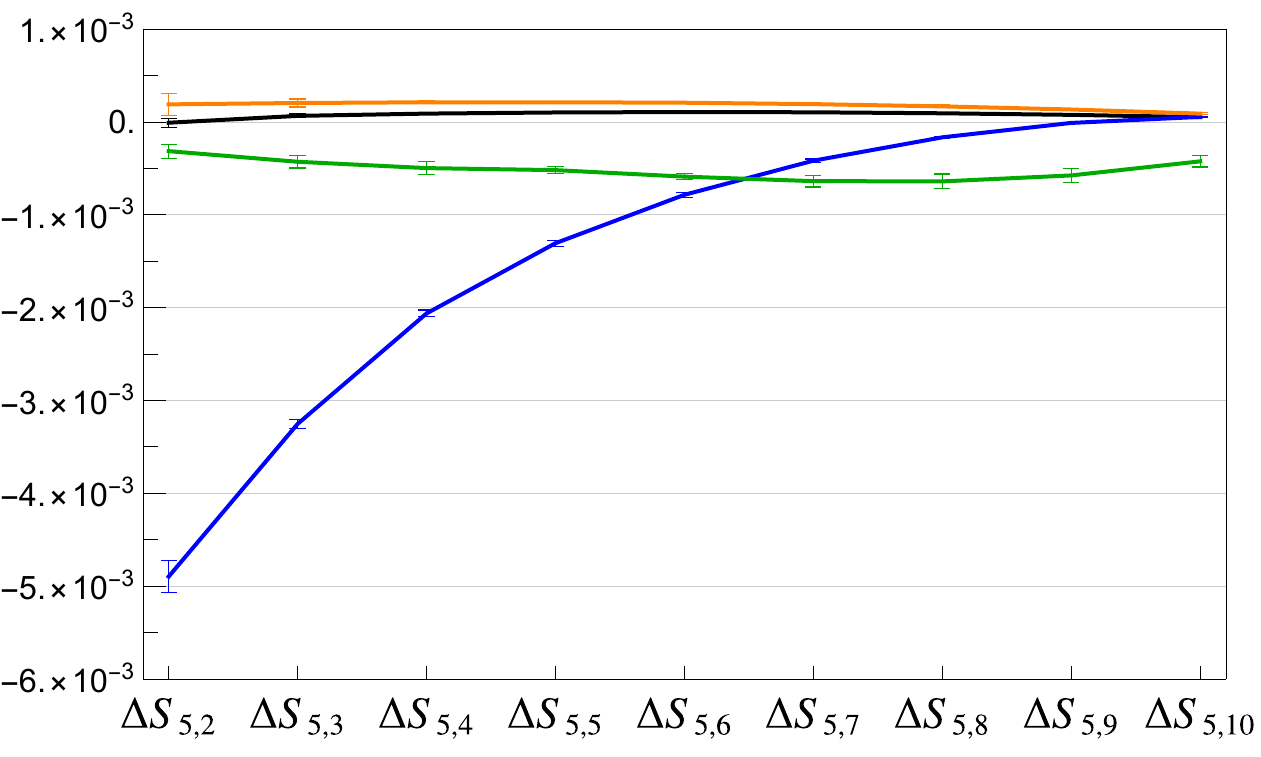}\,
	\caption{$q^2$ distributions of the three angular observables in the lower panel of Table~\ref{tab:obs_meas}.}
	\label{fig:pred2}
\end{figure*}

 We now calculate $A^{D^*}_{\text{FB}}(q^2),  \tilde{F}_L(q^2),  F_L(q^2)$ and $ S_3(q^2)$ for our BP scenarios. The binned observables are defined by
 \beq
 O_i \equiv \frac{1}{\Gamma_{\text{tot}}^{D^{(*)}}}\int_{q^2_{i-1}}^{q^2_{i}}dq^2O(q^2)\Gamma^{D^{(*)}}_f(q^2),\quad i = 2~\text{to}~10\,.
 \eeq
We present the four binned angular observables for the three BPs in Fig.~\ref{fig:obs}. We also show the normalized $q^2$ distribution for $\bar B\ra D\ell\bar X$. 
Large deviations from the SM are evident in several $q^2$ bins.
The error bars in the middle and lower panels indicate the uncertainties due to the hadronic form factors. 
We estimate these as the range of predictions using our chosen form factors~\cite{Bordone:2019guc} and the  form factors of Refs.~\cite{Tanaka:2012nw, Iguro:2020cpg}.  We see that $\Delta S_3$ is quite sensitive to the form factor.

Other observables that have not yet been measured and can be significantly modified by NP include the forward-backward asymmetry in $\bar B\ra D \ell \bar X$,  $A^D_{\text{FB}}$.
In the SM, this is suppressed by $m_{\ell}^2$. In the limit $m_{\ell} \ra 0$, $A^D_{\text{FB}}$ is proportional to $q^2$ with the new LH interactions  $(\mathcal{O}^{S}_{LL}+\mathcal{O}^{S}_{RL}) \mathcal{O}^T_{LL}$. With the new RH interactions $(\mathcal{O}^V_{LR}+\mathcal{O}^V_{RR})^2$, $A^D_{\text{FB}}$  is proportional to $m_N^2 (1-m_N^2/q^2)$,  and for $(\mathcal{O}^{S}_{LR}+\mathcal{O}^{S}_{RR}) \mathcal{O}^{T}_{RR}$, it is proportional to  $q^2 (1-m_N^2/q^2)$.
The $q^2$ averaged values of $\Delta A^D_{\text{FB}}, \Delta S_4$ and  $\Delta S_5$ for the BPs are displayed in Table~\ref{tab:obs_meas}.
In Fig.~\ref{fig:pred2}, we plot the corresponding $q^2$ binned observables and find that large deviations from the SM are possible.



{\bf Summary.} We have presented the angular distributions
for $\bar B\ra D \ell \bar X$ and $\bar B\ra D^{*}  \ell \bar X \ra D\pi  \ell \bar X$, where $X$  may be a massive RH neutrino, for the most general set of operators in LEFT; see section IV of {\it Supplemental Material} for complete expressions. 
 Interestingly, compared to the massless RH neutrino case, no new angular structures result.
However, to obtain a positive value of $\Delta \langle A^{D^*}_{\text{FB}}\rangle$, as suggested by Belle data, a nonzero $m_N$ is needed if the new physics only affects the muon sector. We also made predictions for several angular observables that differ substantially from  SM expectations.

{\bf Acknowledgments.}
We thank P.~Urquijo for a useful discussion.  A.D. is supported in part by the U.S. National Science Foundation under Grant No.~ PHY-1915142. H.L. is supported by ISF, BSF and Azrieli foundation. D.M. is supported in part by the U.S. Department of Energy under Grant No. de-sc0010504. D.M. thanks KITP, Santa Barbara for its hospitality, and support via the NSF under Grant No. PHY-1748958, during the completion of this work.

\newpage
\twocolumngrid
\vspace{-8pt}
\section*{References}
\vspace{-10pt}
\def\bibsection{}
\bibliography{Nref}

\input{supp}

\end{document}

%% file: supp.tex
\clearpage
\newpage
\maketitle
\onecolumngrid

\begin{center}
	\textbf{\large $\bar B \to  D^{(*)} \ell \bar{X}$  decays in effective field theory with massive right-handed neutrinos} \\
	\vspace{0.05in}
	{ \it \large Supplemental Material}\\
	\vspace{0.05in}
	{Alakabha Datta, Hongkai Liu, and Danny Marfatia}
\end{center}

\onecolumngrid
\setcounter{equation}{0}
\setcounter{figure}{0}
\setcounter{table}{0}
\setcounter{section}{0}
\setcounter{page}{1}
\makeatletter
\renewcommand{\theequation}{S\arabic{equation}}
\renewcommand{\thefigure}{S\arabic{figure}}
\renewcommand{\thetable}{S\arabic{table}}
\newcommand\ptwiddle[1]{\mathord{\mathop{#1}\limits^{\scriptscriptstyle(\sim)}}}

\section{Helicity amplitudes}

\label{app:HM}
The helicity amplitude for $\bar B \ra M \ell \bar X$ can be divided into several components according to the helicity  of the $\ell$ ($\lambda_{\ell}$), $\bar X$ ($\lambda_{\bar X}$),  and meson $M$ ($\lambda_M$).
For the $D$ ($D^*$) meson $\lambda_D = s$ ($\lambda_{D^*} = \pm,0$). The total amplitude including the SM contribution is
\beqa
\mathcal{M}(\lambda_M,\lambda_{\ell},\lambda_{\bar X}) &=& \frac{G_F V_{cb}}{\sqrt{2}}\sum_{\alpha=L,R}\sum_{\beta=L,R}[(\sum_{\lambda=0,\pm,t}\delta_{L\alpha} \delta_{L\beta}+C^V_{\alpha\beta})\delta_{\epsilon}(\lambda) H^{V,\alpha}_{\lambda_M,\lambda}L^{V,\beta}_{\lambda_{\ell},\lambda_{\bar X},\lambda} + C^S_{\alpha\beta} H^{S,\alpha}_{\lambda_M} L^{S,\beta}_{\lambda_{\ell},\lambda_{\bar X}}\nonumber\\
&+&  \sum_{\lambda=0,\pm,t}\sum_{\lambda^{\prime} = 0,\pm,t}C^T_{\alpha\beta} \delta_{\epsilon}(\lambda) \delta_{\epsilon}(\lambda^{\prime})H^{T,\alpha}_{\lambda_M,\lambda \lambda^{\prime}} L^{T,\beta}_{\lambda_{\ell},\lambda_{\bar X},\lambda\lambda^{\prime}}]\,,
\label{eq:MDs}
\eeqa
where $L$'s and $H$'s are the leptonic and hadronic amplitudes respectively, which are given in sections~\ref{app:amp_L} and~\ref{app:amp_H}. $\alpha$ and $\beta$ indicate the chirality of the hadronic and leptonic operators, respectively. $\delta_{L\alpha}\delta_{L\beta}$ represents the SM contribution. $\lambda$ and $\lambda^{\prime}$ denote the helicity of virtual vector bosons. In the rest frame, the polarization vectors are
\beq
\epsilon_\mu(0)= \{0,0,0,-1\}\,,\quad \epsilon_\mu(\pm) = \mp\frac{1}{\sqrt{2}} \{0,1,\mp i,0\}\,,\quad \epsilon_\mu(s) = \{1,0,0,0\}\,.
\eeq
The completeness relation is
\beq
\sum_{\lambda}\delta_{\epsilon}(\lambda) \epsilon^*_\mu(\lambda)\epsilon_\nu(\lambda) = g_{\mu\nu}\quad \text{with} \quad \delta_{\epsilon}(0) = \delta_{\epsilon}(\pm)=-\delta_{\epsilon}(s) = -1\,.
\eeq
\section{Leptonic amplitudes}
\label{app:amp_L}
We derive the leptonic amplitudes with massive right-handed neutrinos. The leptonic helicity amplitudes are defined by~\cite{Tanaka:2012nw}
\beqa
L^{V,L/R}_{\lambda_{\ell},\lambda_{\bar X},\lambda} &=& \epsilon_{\mu}(\lambda)\langle\ell(\lambda_{\ell})\bar\nu(\lambda_{\bar X})|\bar\ell\gamma^{\mu}(1\mp\gamma_5)\nu |0\rangle\,,\\
L^{S,L/R}_{\lambda_{\ell},\lambda_{\bar X}} &=& \langle\ell(\lambda_{\ell})\bar\nu(\lambda_{\bar X})|\bar\ell(1\mp\gamma_5)\nu |0\rangle\,,\\		
L^{T,L/R}_{\lambda_{\ell},\lambda_{\bar X},\lambda\lambda^{\prime}} &=& -L^{T,L/R}_{\lambda_{\ell},\lambda_{\bar X},\lambda^{\prime}\lambda} =-i \epsilon_{\mu}(\lambda)\epsilon_{\nu}(\lambda^{\prime})\langle\ell(\lambda_{\ell})\bar\nu(\lambda_{\bar X})|\bar\ell\sigma^{\mu\nu}(1\mp\gamma_5)\nu |0\rangle\,.
\eeqa
We assume that the left-handed neutrinos are the SM neutrinos and right-handed neutrinos are the new sterile neutrinos.  Left-handed operators only couple to the SM left-handed neutrinos or SM right-handed antineutrinos. Thus, by definition
\beq
L^{V,L}_{\lambda_{\ell},-\frac{1}{2},\lambda}=L^{S,L}_{\lambda_{\ell},-\frac{1}{2}}=L^{T,L}_{\lambda_{\ell},-\frac{1}{2},\lambda\lambda^{\prime}}=0\,.
\eeq
The vector leptonic helicity amplitudes with $\beta=L$ and massless right-handed antineutrinos ($\lambda_{\bar X} = 1/2$) are
\beqa
L^{V,L}_{-\frac{1}{2},\frac{1}{2},\lambda_{\epsilon}}&=&\{-\sin\theta_\ell, e^{- i\phi}\frac{1+\cos\theta_\ell}{\sqrt{2}}, e^{ i\phi}\frac{1-\cos\theta_\ell}{\sqrt{2}}, 0\} K_{++}(0)\,,\\
L^{V,L}_{\frac{1}{2},\frac{1}{2},\lambda_{\epsilon}}&=&\{e^{- i\phi}\cos\theta_\ell, e^{-2 i\phi} \frac{\sin\theta_\ell}{\sqrt{2}}, -\frac{\sin\theta_\ell}{\sqrt{2}}, -e^{- i\phi}\}K_{+-}(0)\,,
\eeqa
where $\lambda_{\epsilon} = \{0, +, - , s\}$ and the kinematic functions $K$ carrying the mass dependence are defined by
\beqa
K_{\pm\pm}(m_N) &=& \frac{(E_N+m_N\pm p_{\ell})(E_{\ell}+m_{\ell}\pm p_{\ell})}{\sqrt{(E_{\ell}+m_{\ell})(E_N+m_N)}}\,.
\label{eq:k}
\eeqa
The kinematics relations are
\beqa
E_{\ell} &=& \frac{m_{\ell}^2-m_N^2+q^2}{2\sqrt{q^2}}\,,\\
E_N &=& \frac{m_N^2-m_{\ell}^2+q^2}{2\sqrt{q^2}}\,,\\
p_{\ell} &=& \frac{\sqrt{-2m_N^2(m_{\ell}^2+q^2)+m_N^4+(m_{\ell}^2-q^2)^2}}{2\sqrt{q^2}}\,.
\eeqa
For massless neutrinos
\beq
K_{++}(0) = 2 \sqrt{q^2} \beta_{\ell}\,,\quad K_{+-}(0) = 2 m_{\ell}\beta_{\ell}\,,\quad K_{-+}(0) =K_{--}(0) = 0\,.
\eeq
In the limit $m_{\ell} \ra 0$,
\beqa
K_{++}(m_N) = 2 \sqrt{q^2} \beta_{N}\,,\quad K_{-+}(m_N) = 2 m_{N}\beta_{N}\,,\quad K_{+-}(m_N) =K_{--}(m_N) = 0\,,
\eeqa
where $\beta_{\ell}\equiv \sqrt{1-m_{\ell}^2/q^2}$ and $\beta_{N}\equiv \sqrt{1-m_{N}^2/q^2}$.

The scalar leptonic helicity amplitudes with $\beta=L$ and massless right-handed antineutrinos are
\beqa
L^{S,L}_{\lambda_{\ell},\frac{1}{2}}&=&\{0,-e^{-i\phi}\}K_{++}(0)\,,
\eeqa
where $\lambda_{\ell} = \{-\frac{1}{2},\frac{1}{2}\}$\,. 

The tensor leptonic helicity amplitudes with $\beta=L$ and massless right-handed antineutrinos are
\beqa
L^{T,L}_{-\frac{1}{2},\frac{1}{2},\lambda_{\epsilon}\lambda^{\prime}_{\epsilon}}&=&\{- e^{- i\phi}\frac{1+\cos\theta_\ell}{\sqrt{2}}, e^{ i\phi}\frac{1-\cos\theta_\ell}{\sqrt{2}},\sin\theta_\ell,-\sin\theta_\ell,-e^{- i\phi} \frac{1+\cos\theta_\ell}{\sqrt{2}},-e^{i\phi} \frac{1-\cos\theta_\ell}{\sqrt{2}}\}K_{+-}(0)\,,\nonumber\\
\\
L^{T,L}_{\frac{1}{2},\frac{1}{2},\lambda_{\epsilon}\lambda^{\prime}_{\epsilon}}&=&\{-e^{-2 i\phi}\frac{\sin\theta_\ell}{\sqrt{2}}, -\frac{\sin\theta_\ell}{\sqrt{2}},-e^{-i\phi}\cos\theta_\ell,e^{-i\phi}\cos\theta_\ell,-e^{-2 i\phi}  \frac{\sin\theta_\ell}{\sqrt{2}},\frac{\sin\theta_\ell}{\sqrt{2}}\}K_{++}(0)\,.
\eeqa
where $\lambda_{\epsilon}\lambda^{\prime}_{\epsilon} = \{0+, 0-, 0s , +-,+s,-s\}$.

For a right-handed leptonic operator ($\beta = R$), a massive left-handed sterile antineutrino with helicity $\lambda_{\bar X} = \pm 1/2$ can be produced. However,  the production of the $\lambda_{\bar X} = 1/2$ state is helicity suppressed.
The vector leptonic helicity amplitudes with $\beta = R$ and massive left-handed antineutrinos are
\beqa
L^{V,R}_{-\frac{1}{2},-\frac{1}{2},\lambda_{\epsilon}}&=&\{e^{i\phi}\cos\theta,\frac{\sin\theta}{\sqrt{2}},- e^{2 i\phi}\frac{\sin\theta}{\sqrt{2}},-e^{ i\phi}\} K_{+-}(m_N)\,,\\
L^{V,R}_{\frac{1}{2},-\frac{1}{2},\lambda_{\epsilon}}&=&\{\sin\theta,e^{- i\phi}\frac{1-\cos\theta}{\sqrt{2}},e^{ i\phi}\frac{1+\cos\theta}{\sqrt{2}}, 0\} K_{++}(m_N)\,,\\
L^{V,R}_{-\frac{1}{2},\frac{1}{2},\lambda_{\epsilon}}&=&\{-\sin\theta, e^{- i\phi}\frac{1+\cos\theta}{\sqrt{2}},e^{ i\phi}\frac{1-\cos\theta}{\sqrt{2}},0\} K_{--}(m_N)\,,\\
L^{V,R}_{\frac{1}{2},\frac{1}{2},\lambda_{\epsilon}}&=&\{e^{- i\phi}\cos\theta, e^{- 2i\phi}\frac{\sin\theta}{\sqrt{2}},-\frac{\sin\theta}{\sqrt{2}},e^{-i\phi}\} K_{-+}(m_N)\,.
\eeqa

The scalar leptonic helicity amplitudes with $\beta=R$ and massive left-handed antineutrinos are
\beqa
L^{S,R}_{\lambda_{\ell},-\frac{1}{2}}&=&\{-e^{i\phi},0\}K_{++}(m_N)\,,\\
L^{S,R}_{\lambda_{\ell},\frac{1}{2}}&=& \{0,e^{-i\phi}\}K_{--}(m_N)\,.
\eeqa

The tensor leptonic helicity amplitudes with $\beta=R$ and massive left-handed antineutrinos are
\beqa
L^{T,R}_{-\frac{1}{2},-\frac{1}{2},\lambda_{\epsilon}\lambda^{\prime}_{\epsilon}}&=&\{\frac{\sin\theta}{\sqrt{2}}, e^{2 i\phi}\frac{\sin\theta}{\sqrt{2}},-e^{i\phi}\cos\theta,-e^{i\phi}\cos\theta,-\frac{\sin\theta}{\sqrt{2}}, e^{2 i\phi}\frac{\sin\theta}{\sqrt{2}}\}K_{++}(m_N)\,,\\
L^{T,R}_{\frac{1}{2},-\frac{1}{2},\lambda_{\epsilon}\lambda^{\prime}_{\epsilon}}&=& \{e^{- i\phi }\frac{1-\cos\theta}{\sqrt{2}},-e^{i\phi }\frac{1+\cos\theta}{\sqrt{2}},-\sin\theta,-\sin\theta,- e^{- i\phi} \frac{1-\cos\theta}{\sqrt{2}},- e^{i\phi} \frac{1+\cos\theta}{\sqrt{2}} \}K_{+-}(m_N)\,,\nonumber\\
\\
L^{T,R}_{-\frac{1}{2},\frac{1}{2},\lambda_{\epsilon}\lambda^{\prime}_{\epsilon}}&=& \{ e^{- i\phi}\frac{1+\cos\theta}{\sqrt{2}},- e^{i\phi}\frac{1-\cos\theta}{\sqrt{2}},\sin\theta,\sin\theta,-e^{- i\phi}\frac{1+\cos\theta}{\sqrt{2}},-e^{ i\phi}\frac{1-\cos\theta}{\sqrt{2}}\}K_{-+}(m_N)\,,\nonumber\\
\\
L^{T,R}_{\frac{1}{2},\frac{1}{2},\lambda_{\epsilon}\lambda^{\prime}_{\epsilon}}&=&\{e^{-2 i\phi }\frac{\sin\theta}{\sqrt{2}},\frac{\sin\theta}{\sqrt{2}},-e^{- i\phi }\cos\theta,-e^{- i\phi }\cos\theta ,-e^{-2 i\phi }\frac{\sin\theta}{\sqrt{2}},\frac{\sin\theta}{\sqrt{2}}\}K_{--}(m_N)\,.
\eeqa
In the massless limit, $m_N \ra 0$, all the terms with $\beta = R$ and $\lambda_{\bar X} = 1/2$ vanish, as there is no helicity flip.

\section{Hadronic amplitudes}
\label{app:amp_H}
For completeness, we provide the hadronic helicity amplitudes available in the literature~\cite{Tanaka:2012nw}:
\beqa
H^{V,L/R}_{\lambda_M,\lambda} &=& \epsilon^*_{\mu}(\lambda)\langle M (\lambda_M)|\bar{c} \gamma^{\mu} (1\mp \gamma_5)b|\bar B\rangle\,,\\
H^{S,L/R}_{\lambda_M} &=& \langle M (\lambda_M)|\bar{c}  (1\mp \gamma_5)b|\bar B\rangle\,,\\
H^{T,L/R}_{\lambda_M,\lambda\lambda^{\prime}} &=& i \epsilon^*_{\mu}(\lambda)\epsilon^*_{\nu} (\lambda^{\prime})\langle M (\lambda_M)|\bar{c} \sigma^{\mu\nu} (1\mp \gamma_5)b|\bar B\rangle\,.
\eeqa 
For $\bar B \ra D\ell\bar X$, the vector amplitudes are
\beqa
H^{V,L/R}_{s,0} &=& H^V_{s,0}(q^2)\,, \quad H^{V,L/R}_{s,s} = H^V_{s,s}(q^2)\,.
\eeqa
And the vector amplitudes for $\bar B \ra D^*\ell\bar X$ are
\beqa
H^{V,L}_{\pm,\pm} = - H^{V,R}_{\mp,\mp} &=& H^V_{\pm}(q^2)\,,\\
H^{V,L}_{0,0} = - H^{V,R}_{0,0} &=& H^{V}_{0}(q^2)\,,\\
H^{V,L}_{0,s} = - H^{V,R}_{0,s}  &=& H^{V}_{s}(q^2)\,. 
\eeqa
The amplitudes of the scalar type operators are defined by
\beqa
H^{S,L/R}_{s} &=& H^{S}_{s}(q^2)\,,\\
H^{S,L/R}_{0}  &=& \mp H^{S}_{0}(q^2)\,.
\eeqa
The amplitudes of the tensor type operators are defined by
\beqa
H^{T,L}_{s,+-}=H^{T,L}_{s,0s}=-H^{T,R}_{s,+-}=H^{T,R}_{s,0s} &=& H^{T}_{s}(q^2),\\
H^{T,L}_{\pm,\pm0}=\pm H^{T,L}_{\pm,\pm s} =- H^{T,R}_{\mp,\mp0} = \mp H^{T,R}_{\mp,\mp s} &=& H^{T}_{\pm}(q^2)\,,\\
H^{T,L}_{0,+-}= H^{T,L}_{0,0s} = H^{T,R}_{0,+-} = -H^{T,R}_{0,0s} &=& H^{T}_{0}(q^2)\,,\\
H^{T,L/R}_{\lambda_M,\lambda\lambda^{\prime}}(q^2) = - H^{T,L/R}_{\lambda_M,\lambda^{\prime}\lambda}(q^2)\,. &\quad&
\eeqa
The hadronic amplitudes including mass corrections up to ${\cal{O}}(1/m_c^2)$ are given in Ref.~\cite{Bordone:2019guc,Iguro:2020cpg}.

\section{Angular distributions with massive right-handed neutrinos} 
\label{app:ang}
The helicity amplitudes (Eq.~\ref{eq:MDs}) of the process $\bar B\ra D \ell \bar{X}$ can be written in terms of 
$\mathcal{\tilde{A}}$ functions as
\beqa
M_D(++) &\equiv& \mathcal{M}(s,+,+)=\mathcal{\tilde{A}}_1^{++}+ \mathcal{\tilde{A}}_2^{++}\cos\theta_{\ell}\,,\\
M_D(-+) &\equiv& \mathcal{M}(s,-,+) =\mathcal{\tilde{A}}^{-+}\sin\theta_{\ell}\,,\\
M_D(+-) &\equiv& \mathcal{M}(s,+,-) =\mathcal{\tilde{A}}^{+-}\sin\theta_{\ell}\,,\\
M_D(--) &\equiv& \mathcal{M}(s,-,-) =\mathcal{\tilde{A}}_1^{--}+ \mathcal{\tilde{A}}_2^{--}\cos\theta_{\ell}\,.
\eeqa
The helicity amplitudes of the process $\bar B\ra \ell \bar{X} D^* (\ra D\pi)$ after summing over the helicity of 
$D^*$ can be written in terms of 
$\mathcal{{A}}$ functions as
\beqa
M_{D^*}(++) &=& \cos\theta_D e^{-i\phi}[\mathcal{A}_1^{++} +\mathcal{A}_2^{++}\cos\theta_{\ell} ] + \sin\theta_D\sin\theta_{\ell}[\mathcal{A}_3^{++}+ \mathcal{A}_4^{++}e^{-2 i\phi} ]\,,\\
M_{D^*}(-+) &=& \mathcal{A}_1^{-+}\cos\theta_D\sin\theta_{\ell} + \sin\theta_D[\mathcal{A}_2^{-+}(1-\cos\theta_{\ell})e^{i\phi} + \mathcal{A}_3^{-+} (1+\cos\theta_{\ell})e^{-i\phi} ]\,,\\
M_{D^*}(+-) &=& \mathcal{A}_1^{+-}\cos\theta_D\sin\theta_{\ell} +  \sin\theta_D[\mathcal{A}_2^{+-} (1- \cos\theta_{\ell}) e^{-i\phi} + \mathcal{A}_3^{+-} (1+\cos\theta_{\ell})e^{i\phi}]\,,\\
M_{D^*}(--) &=& \cos\theta_D e^{i\phi}[\mathcal{A}_1^{--} +\mathcal{A}_2^{--}\cos\theta_{\ell} ] + \sin\theta_D\sin\theta_{\ell}[\mathcal{A}_3^{--}+ \mathcal{A}_4^{--}e^{2 i\phi} ]\,.
\eeqa
Note that $\mathcal{\tilde{A}}$ and $\mathcal{A}$ are functions of the Wilson coefficients $C$ defined in  Eq.~(2), the hadronic amplitudes $H$ given in section~\ref{app:amp_H}, and the kinematic variables $K$ defined in Eq.~(\ref{eq:k}). For massless LH neutrinos ($\beta = L$),

\beqa
\mathcal{\tilde{A}}_1^{++} &=& -(C^V_{RL}  + C^V_{LL} )H^V_{s,s} K_{+-}(0)  - (C^S_{LL} +C^S_{RL}) H^S_s K_{++}(0)\,,\\
\mathcal{\tilde{A}}_2^{++} &=& - (C^V_{LL}  + C^V_{RL} )H^V_{s,0} K_{+-}(0)+ 4 C^T_{LL} H^{T}_s K_{++}(0)\,,\\
\mathcal{\tilde{A}}^{-+} &=&  (C^V_{LL} + C^V_{RL}) H^V_{s,0} K_{++}(0) - 4 C^T_{LL} H^T_s K_{+-}(0)\,,\\
\mathcal{\tilde{A}}^{+-} &=& \mathcal{\tilde{A}}_1^{--} = \mathcal{\tilde{A}}_2^{--} = 0\,,
\eeqa

\beqa
\mathcal{A}_1^{++} &=& (C^S_{RL}-C^S_{LL})H^{S}_0 K_{++}(0)
+(C^V_{LL} - C^V_{RL})H^V_{s} K_{+-}(0)\,,\\
\mathcal{A}_2^{++} &=&   (C^V_{LL}  - C^V_{RL} )H^V_{0}K_{+-}(0)   - 4 C^T_{LL} H^T_{0} K_{++}(0)\,,\\
\mathcal{A}_3^{++} &=&   \frac{1}{2}(C^V_{RL} H^V_{+} -C^V_{LL} H^V_{-} ) K_{+-}(0)-2 C^T_{LL} H^T_{-} K_{++}(0)\,,\\
\mathcal{A}_4^{++} &=& \frac{1}{2}(C^V_{RL} H^V_{-}  - C^V_{LL} H^V_{+} ) K_{+-}(0) +2 C^T_{LL} H^T_{+} K_{++}(0)\,,
\eeqa

\beqa
\mathcal{A}_1^{-+} &=&  ( C^V_{RL}-C^V_{LL} ) H^V_{0} K_{++}(0)  + 4 C^T_{LL} H^T_{0} K_{+-}(0)\,,\\
\mathcal{A}_2^{-+} &=&   \frac{1}{2}(C^V_{LL} H^V_{-} -C^V_{RL} H^V_{+} ) K_{++}(0)+2 C^T_{LL} H^T_{-} K_{+-}(0)\,,\\
\mathcal{A}_3^{-+} &=&  \frac{1}{2}(C^V_{RL} H^V_{-} - C^V_{LL} H^V_{+})K_{++}(0)+2 C^T_{LL} H^T_{+} K_{+-}(0)\,
%
\eeqa

\beqa
\mathcal{A}_1^{+-} &=& \mathcal{A}_2^{+-}  = \mathcal{A}_3^{+-} = 0\,,
\eeqa

\beqa
\mathcal{A}_1^{--} &=& \mathcal{A}_2^{--}  = \mathcal{A}_3^{--} =\mathcal{A}_4^{--} = 0\,.
\eeqa
With only massless LH neutrinos, the helicity amplitudes $M_{D^{(*)}}(+-) $ and $ M_{D^{(*)}}(--) $ are always zero.
For massive RH neutrinos ($\beta = R$),

\beqa
\mathcal{\tilde{A}}_1^{++} &=& (C^V_{LR} + C^V_{RR} )H^V_{s,s} K_{-+}(m_N) +(C^S_{LR} +C^S_{RR}) H^S_s K_{--}(m_N)\,,\\
\mathcal{\tilde{A}}_2^{++} &=& -(C^V_{LR} + C^V_{RR} )H^V_{s,0} K_{-+}(m_N) + 4 C^T_{RR} K_{--} (m_N)  H^{T}_s\,,\\
\mathcal{\tilde{A}}^{-+} &=& (C^V_{LR} + C^V_{RR})H^V_{s,0} K_{--}(m_N) - 4 C^T_{RR} H^T_s K_{-+}(m_N)\,,\\
\mathcal{\tilde{A}}^{+-} &=& -(C^V_{LR} + C^V_{RR}) H^V_{s,0}K_{++}(m_N) + 4 C^T_{RR} H^T_s K_{+-}(m_N)\,,\\
\mathcal{\tilde{A}}_1^{--} &=& -(C^V_{LR} + C^V_{RR}) H^V_{s,s}K_{+-}(m_N) - (C^S_{LR} + C^S_{RR})H^S_s K_{++}(m_N)\,,\\
\mathcal{\tilde{A}}_2^{--} &=& -(C^V_{LR} + C^V_{RR}) H^V_{s,0}K_{+-}(m_N) + 4 C^T_{RR} H^T_s K_{++}(m_N)\,,
\eeqa

\beqa
\mathcal{A}_1^{++} &=& (C^S_{LR}-C^S_{RR})H^S_{0}K_{--}(m_N)+(C^V_{RR}-C^V_{LR})H^V_{s}K_{-+}(m_N)\,,\\
\mathcal{A}_2^{++} &=&  (C^V_{LR}  - C^V_{RR} )H^V_{0} K_{-+}(m_N) + 4 C^T_{RR} H^T_{0} K_{--}(m_N)\,,\\
\mathcal{A}_3^{++} &=& \frac{1}{2}(C^V_{RR} H^V_{+} -C^V_{LR} H^V_{-} ) K_{-+}(m_N)-2 C^T_{RR} H^T_{+} K_{--}(m_N)\,,\\
\mathcal{A}_4^{++} &=& \frac{1}{2}(C^V_{RR} H^V_{-} - C^V_{LR} H^V_{+})K_{-+}(m_N) +2 C^T_{RR} H^T_{-} K_{--}(m_N)\,,
\eeqa

\beqa
\mathcal{A}_1^{-+} &=& ( C^V_{RR} -C^V_{LR}  )H^V_{0} K_{--}(m_N) - 4 C^T_{RR} H^T_{0} K_{-+}(m_N)\,,\\
\mathcal{A}_2^{-+} &=& \frac{1}{2}(C^V_{LR} H^V_{-} -C^V_{RR} H^V_{+} ) K_{--}(m_N) + 2 C^T_{RR} H^T_{+} K_{-+}(m_N)\,,\\
\mathcal{A}_3^{-+} &=& \frac{1}{2}(C^V_{RR} H^V_{-} - C^V_{LR} H^V_{+})K_{--}(m_N)  +2 C^T_{RR} H^T_{-} K_{-+}(m_N)\,,
%
\eeqa

\beqa
\mathcal{A}_1^{+-} &=& ( C^V_{LR} -C^V_{RR}  )H^V_{0} K_{++}(m_N) + 4 C^T_{RR} H^T_{0}K_{+-}(m_N)\,, \\
\mathcal{A}_2^{+-} &=& \frac{1}{2}(C^V_{RR} H^V_{-} - C^V_{LR} H^V_{+})K_{++}(m_N) + 2 C^T_{RR} H^T_{-} K_{+-}(m_N)\,, \\
\mathcal{A}_3^{+-} &=& \frac{1}{2}(C^V_{LR} H^V_{-} -C^V_{RR} H^V_{+} ) K_{++}(m_N)+ 2 C^T_{RR} H^T_{+} K_{+-}(m_N)\,,
%
\eeqa

\beqa
\mathcal{A}_1^{--} &=& ( C^V_{LR} -C^V_{RR}  )H^V_{s} K_{+-}(m_N) + (C^S_{RR}-C^S_{LR}) H^S_{0}K_{++}(m_N)\,, \\
\mathcal{A}_2^{--} &=& (C^V_{LR}- C^V_{RR} )H^V_{0}K_{+-}(m_N) + 4 C^T_{RR} H^T_{0} K_{++}(m_N)\,, \\
\mathcal{A}_3^{--} &=& \frac{1}{2}(C^V_{RR} H^V_{-} -C^V_{LR} H^V_{+} ) K_{+-}(m_N)+ 
2 C^T_{RR} H^T_{-} K_{++}(m_N)\,,\\
\mathcal{A}_4^{--} &=& \frac{1}{2}(C^V_{RR} H^V_{+} - C^V_{LR} H^V_{-})K_{+-}(m_N) 
-2 C^T_{RR} H^T_{+} K_{++}(m_N)\,.
\eeqa
As expected, the helicity amplitudes $M_{D^{(*)}}(-+)$ and $M_{D^{(*)}}(++)$ from massive RH neutrinos are only related to $K_{--}(m_N)$ and $K_{-+}(m_N)$, which are suppressed by $m_N$.

The differential decay width of  $\bar B\ra D \ell \bar{X} $ can be decomposed into three angular terms: 
\beqa
\frac{d^2\Gamma_{D}}{dq^2d\cos\theta_{\ell}} &=& \mathcal{J}_0(q^2) + \mathcal{J}_1(q^2)\cos\theta_{\ell} + \mathcal{J}_2(q^2)\cos^2\theta_{\ell}\,,
\eeqa
where the three $\mathcal{J}$ coefficients can be written as functions of $\mathcal{\tilde{A}}$:
\beqa
\mathcal{J}_0/\tilde{\mathcal{N}} &=& |\mathcal{\tilde{A}}_1^{--}|^2+|\mathcal{\tilde{A}}_1^{++}|^2+|\mathcal{\tilde{A}}^{+-}|^2+|\mathcal{\tilde{A}}^{-+}|^2\,,\\
\mathcal{J}_1/\tilde{\mathcal{N}}  &=& 2\text{Re}[\mathcal{\tilde{A}}_1^{--}\mathcal{\tilde{A}}_2^{--*} + \mathcal{\tilde{A}}_1^{++}\mathcal{\tilde{A}}_2^{++*}]\,,\\
\mathcal{J}_2/\tilde{\mathcal{N}}  &=&|\mathcal{\tilde{A}}_2^{--}|^2+|\mathcal{\tilde{A}}_2^{++}|^2-|\mathcal{\tilde{A}}^{-+}|^2-|\mathcal{\tilde{A}}^{+-}|^2\,,
\eeqa
where $\tilde{\mathcal{N}} = \frac{G_F^2 V_{cb}^2}{1024\pi^3}\frac{\lambda^{1/2}(m_B^2,m_D^2,q^2)\lambda^{1/2} (q^2,m_{\ell}^2,m_N^2) }{m_B^3q^2}$.

For the decay process $\bar B\ra D^* \ell \bar{X} \ra D\pi\ell\bar{X}$, there are additional angular structures because of the four-body  final state.
The differential decay width can be decomposed into 12 terms:
\beqa
\frac{8\pi}{3} \frac{d^4\Gamma_{D^*}}{dq^2d\cos\theta_{\ell}d\cos\theta_Dd\phi} &=& (\mathcal{I}_{1s}+\mathcal{I}_{2s}\cos2\theta_{\ell}+ \mathcal{I}_{6s}\cos\theta_{\ell})\sin^2\theta_D \nonumber\\
&+&(\mathcal{I}_{1c}+\mathcal{I}_{2c}\cos2\theta_{\ell}+ \mathcal{I}_{6c}\cos\theta_{\ell})\cos^2\theta_D \nonumber\\
&+&(\mathcal{I}_3\cos2\phi+\mathcal{I}_9\sin2\phi )\sin^2\theta_D \sin^2\theta_{\ell}\\
&+&(\mathcal{I}_4\cos\phi+\mathcal{I}_8\sin\phi )\sin2\theta_D \sin2\theta_{\ell}\nonumber\\
&+&(\mathcal{I}_5\cos\phi+\mathcal{I}_7\sin\phi )\sin2\theta_D \sin\theta_{\ell}\nonumber\,,
\eeqa
where the 12 $\mathcal{I}$ coefficients can be written in terms of the $\mathcal{A}$ functions:
\beqa
\mathcal{I}_{1s}/\mathcal{N} &=& \frac{1}{2}(|\mathcal{A}_3^{--}|^2+|\mathcal{A}_3^{++}|^2+|\mathcal{A}_4^{--}|^2+|\mathcal{A}_4^{++}|^2)\nonumber\\
&+&\frac{3}{2}(|\mathcal{A}_3^{-+}|^2+|\mathcal{A}_3^{+-}|^2+|\mathcal{A}_2^{-+}|^2 + |\mathcal{A}_2^{+-}|^2)\,,\\
\mathcal{I}_{1c}/\mathcal{N}&=& \frac{1}{2}(|\mathcal{A}_2^{--}|^2 + |\mathcal{A}_1^{-+}|^2+|\mathcal{A}_1^{+-}|^2+|\mathcal{A}_2^{++}|^2)+|\mathcal{A}_1^{--}|^2+|\mathcal{A}_1^{++}|^2\,,\\
\mathcal{I}_{2s}/\mathcal{N} &=&\frac{1}{2}(|\mathcal{A}_3^{-+}|^2+|\mathcal{A}_2^{-+}|^2+|\mathcal{A}_3^{+-}|^2+|\mathcal{A}_2^{+-}|^2)\nonumber\\
&-&\frac{1}{2}(|\mathcal{A}_3^{--}|^2+|\mathcal{A}_4^{--}|^2+|\mathcal{A}_3^{++}|^2+|\mathcal{A}_4^{++}|^2)\,,\\
\mathcal{I}_{2c}/\mathcal{N}&=&\frac{1}{2}(|\mathcal{A}_2^{--}|^2 + |\mathcal{A}_2^{++}|^2 - |\mathcal{A}_1^{+-}|^2-|\mathcal{A}_1^{-+}|^2)\,,\\
\mathcal{I}_3/\mathcal{N}&=&2\text{Re}[\mathcal{A}_3^{--}\mathcal{A}_4^{--*} + \mathcal{A}_3^{++}\mathcal{A}_4^{++*}+\mathcal{A}_3^{-+}\mathcal{A}_2^{-+*}+\mathcal{A}_3^{+-}\mathcal{A}_2^{+-*}]\,,\\
\mathcal{I}_4/\mathcal{N}&=&\frac{1}{2}\text{Re}[\mathcal{A}_2^{--}(\mathcal{A}_3^{--*}+\mathcal{A}_4^{--*}) + \mathcal{A}_1^{-+}(\mathcal{A}_3^{-+*}-\mathcal{A}_2^{-+*}) 
+ \mathcal{A}_1^{+-}(\mathcal{A}_3^{+-*}-\mathcal{A}_2^{+-*})\nonumber\\ &\quad&+\mathcal{A}_2^{++}(\mathcal{A}_3^{++*}+\mathcal{A}_4^{++*}) ]\,,\\
\mathcal{I}_5/\mathcal{N}&=&\text{Re}[\mathcal{A}_1^{--}(\mathcal{A}_3^{--*}+\mathcal{A}_4^{--*})+\mathcal{A}_1^{-+}(\mathcal{A}_2^{-+*}+\mathcal{A}_3^{-+*}) +\mathcal{A}_1^{+-}(\mathcal{A}_2^{+-*}+\mathcal{A}_3^{+-*}) \nonumber\\ &\quad&+\mathcal{A}_1^{++}(\mathcal{A}_3^{++*}+\mathcal{A}_4^{++*})  ]\,,\\
\mathcal{I}_{6s}/\mathcal{N}&=& 2(|\mathcal{A}_3^{-+}|^2+|\mathcal{A}_3^{+-}|^2 - |\mathcal{A}_2^{-+}|^2 - |\mathcal{A}_2^{+-}|^2)\,, \\
\mathcal{I}_{6c}/\mathcal{N}&=&2\text{Re}[\mathcal{A}_1^{--}\mathcal{A}_2^{--*} + \mathcal{A}_1^{++}\mathcal{A}_2^{++*}]\,, \\
\mathcal{I}_7/\mathcal{N}&=& \text{Im}[\mathcal{A}_1^{++}(\mathcal{A}_3^{++*}-\mathcal{A}_4^{++*}) + \mathcal{A}_1^{-+}(\mathcal{A}_2^{-+*}-\mathcal{A}_3^{-+*})\nonumber\\
&\quad&- \mathcal{A}_1^{+-}(\mathcal{A}_2^{+-*}-\mathcal{A}_3^{+-*})-\mathcal{A}_1^{--}(\mathcal{A}_3^{--*}-\mathcal{A}_4^{--*})]\,, \\
\mathcal{I}_8/\mathcal{N}&=&\frac{1}{2}\text{Im}[\mathcal{A}_1^{+-}(\mathcal{A}_3^{+-*}+\mathcal{A}_2^{+-*})+\mathcal{A}_2^{++}(\mathcal{A}_3^{++*}-\mathcal{A}_4^{++*}) \nonumber\\
&\quad&- \mathcal{A}_1^{-+}(\mathcal{A}_3^{-+*}+\mathcal{A}_2^{-+*})-\mathcal{A}_2^{--}(\mathcal{A}_3^{--*}-\mathcal{A}_4^{--*})]\,, \\
\mathcal{I}_9/\mathcal{N}&=& 2 \text{Im}[\mathcal{A}_3^{--}\mathcal{A}_4^{--*}+\mathcal{A}_3^{-+}\mathcal{A}_2^{-+*}-\mathcal{A}_3^{++}\mathcal{A}_4^{++*}-\mathcal{A}_3^{+-}\mathcal{A}_2^{+-*}]\,,
\eeqa
where $\mathcal{N} = \frac{G_F^2 V_{cb}^2}{512\pi^3}\frac{\lambda^{1/2}(m_B^2,m_{D^*}^2,q^2)\lambda^{1/2} (q^2,m_{\ell}^2,m_N^2) }{m_B^3q^2}~\cB(D^*\ra D\pi)$.

%% file: main.bbl
\begin{thebibliography}{40}%
\makeatletter
\providecommand \@ifxundefined [1]{%
 \@ifx{#1\undefined}
}%
\providecommand \@ifnum [1]{%
 \ifnum #1\expandafter \@firstoftwo
 \else \expandafter \@secondoftwo
 \fi
}%
\providecommand \@ifx [1]{%
 \ifx #1\expandafter \@firstoftwo
 \else \expandafter \@secondoftwo
 \fi
}%
\providecommand \natexlab [1]{#1}%
\providecommand \enquote  [1]{``#1''}%
\providecommand \bibnamefont  [1]{#1}%
\providecommand \bibfnamefont [1]{#1}%
\providecommand \citenamefont [1]{#1}%
\providecommand \href@noop [0]{\@secondoftwo}%
\providecommand \href [0]{\begingroup \@sanitize@url \@href}%
\providecommand \@href[1]{\@@startlink{#1}\@@href}%
\providecommand \@@href[1]{\endgroup#1\@@endlink}%
\providecommand \@sanitize@url [0]{\catcode `\\12\catcode `\$12\catcode
  `\&12\catcode `\#12\catcode `\^12\catcode `\_12\catcode `\%12\relax}%
\providecommand \@@startlink[1]{}%
\providecommand \@@endlink[0]{}%
\providecommand \url  [0]{\begingroup\@sanitize@url \@url }%
\providecommand \@url [1]{\endgroup\@href {#1}{\urlprefix }}%
\providecommand \urlprefix  [0]{URL }%
\providecommand \Eprint [0]{\href }%
\providecommand \doibase [0]{http://dx.doi.org/}%
\providecommand \selectlanguage [0]{\@gobble}%
\providecommand \bibinfo  [0]{\@secondoftwo}%
\providecommand \bibfield  [0]{\@secondoftwo}%
\providecommand \translation [1]{[#1]}%
\providecommand \BibitemOpen [0]{}%
\providecommand \bibitemStop [0]{}%
\providecommand \bibitemNoStop [0]{.\EOS\space}%
\providecommand \EOS [0]{\spacefactor3000\relax}%
\providecommand \BibitemShut  [1]{\csname bibitem#1\endcsname}%
\let\auto@bib@innerbib\@empty
\bibitem [{\citenamefont {Lees}\ \emph {et~al.}(2012)\citenamefont {Lees} \emph
  {et~al.}}]{BaBar:2012obs}%
  \BibitemOpen
  \bibfield  {author} {\bibinfo {author} {\bibfnamefont {J.~P.}\ \bibnamefont
  {Lees}} \emph {et~al.} (\bibinfo {collaboration} {BaBar}),\ }\href {\doibase
  10.1103/PhysRevLett.109.101802} {\bibfield  {journal} {\bibinfo  {journal}
  {Phys. Rev. Lett.}\ }\textbf {\bibinfo {volume} {109}},\ \bibinfo {pages}
  {101802} (\bibinfo {year} {2012})},\ \Eprint {http://arxiv.org/abs/1205.5442}
  {arXiv:1205.5442 [hep-ex]} \BibitemShut {NoStop}%
\bibitem [{\citenamefont {Lees}\ \emph {et~al.}(2013)\citenamefont {Lees} \emph
  {et~al.}}]{BaBar:2013mob}%
  \BibitemOpen
  \bibfield  {author} {\bibinfo {author} {\bibfnamefont {J.~P.}\ \bibnamefont
  {Lees}} \emph {et~al.} (\bibinfo {collaboration} {BaBar}),\ }\href {\doibase
  10.1103/PhysRevD.88.072012} {\bibfield  {journal} {\bibinfo  {journal} {Phys.
  Rev. D}\ }\textbf {\bibinfo {volume} {88}},\ \bibinfo {pages} {072012}
  (\bibinfo {year} {2013})},\ \Eprint {http://arxiv.org/abs/1303.0571}
  {arXiv:1303.0571 [hep-ex]} \BibitemShut {NoStop}%
\bibitem [{\citenamefont {Aaij}\ \emph {et~al.}(2015)\citenamefont {Aaij} \emph
  {et~al.}}]{LHCb:2015gmp}%
  \BibitemOpen
  \bibfield  {author} {\bibinfo {author} {\bibfnamefont {R.}~\bibnamefont
  {Aaij}} \emph {et~al.} (\bibinfo {collaboration} {LHCb}),\ }\href {\doibase
  10.1103/PhysRevLett.115.111803} {\bibfield  {journal} {\bibinfo  {journal}
  {Phys. Rev. Lett.}\ }\textbf {\bibinfo {volume} {115}},\ \bibinfo {pages}
  {111803} (\bibinfo {year} {2015})},\ \bibinfo {note} {[Erratum:
  Phys.Rev.Lett. 115, 159901 (2015)]},\ \Eprint
  {http://arxiv.org/abs/1506.08614} {arXiv:1506.08614 [hep-ex]} \BibitemShut
  {NoStop}%
\bibitem [{\citenamefont {Huschle}\ \emph {et~al.}(2015)\citenamefont {Huschle}
  \emph {et~al.}}]{Belle:2015qfa}%
  \BibitemOpen
  \bibfield  {author} {\bibinfo {author} {\bibfnamefont {M.}~\bibnamefont
  {Huschle}} \emph {et~al.} (\bibinfo {collaboration} {Belle}),\ }\href
  {\doibase 10.1103/PhysRevD.92.072014} {\bibfield  {journal} {\bibinfo
  {journal} {Phys. Rev. D}\ }\textbf {\bibinfo {volume} {92}},\ \bibinfo
  {pages} {072014} (\bibinfo {year} {2015})},\ \Eprint
  {http://arxiv.org/abs/1507.03233} {arXiv:1507.03233 [hep-ex]} \BibitemShut
  {NoStop}%
\bibitem [{\citenamefont {Sato}\ \emph {et~al.}(2016)\citenamefont {Sato} \emph
  {et~al.}}]{Belle:2016ure}%
  \BibitemOpen
  \bibfield  {author} {\bibinfo {author} {\bibfnamefont {Y.}~\bibnamefont
  {Sato}} \emph {et~al.} (\bibinfo {collaboration} {Belle}),\ }\href {\doibase
  10.1103/PhysRevD.94.072007} {\bibfield  {journal} {\bibinfo  {journal} {Phys.
  Rev. D}\ }\textbf {\bibinfo {volume} {94}},\ \bibinfo {pages} {072007}
  (\bibinfo {year} {2016})},\ \Eprint {http://arxiv.org/abs/1607.07923}
  {arXiv:1607.07923 [hep-ex]} \BibitemShut {NoStop}%
\bibitem [{\citenamefont {Hirose}\ \emph {et~al.}(2017)\citenamefont {Hirose}
  \emph {et~al.}}]{Belle:2016dyj}%
  \BibitemOpen
  \bibfield  {author} {\bibinfo {author} {\bibfnamefont {S.}~\bibnamefont
  {Hirose}} \emph {et~al.} (\bibinfo {collaboration} {Belle}),\ }\href
  {\doibase 10.1103/PhysRevLett.118.211801} {\bibfield  {journal} {\bibinfo
  {journal} {Phys. Rev. Lett.}\ }\textbf {\bibinfo {volume} {118}},\ \bibinfo
  {pages} {211801} (\bibinfo {year} {2017})},\ \Eprint
  {http://arxiv.org/abs/1612.00529} {arXiv:1612.00529 [hep-ex]} \BibitemShut
  {NoStop}%
\bibitem [{\citenamefont {Aaij}\ \emph
  {et~al.}(2018{\natexlab{a}})\citenamefont {Aaij} \emph
  {et~al.}}]{LHCb:2017smo}%
  \BibitemOpen
  \bibfield  {author} {\bibinfo {author} {\bibfnamefont {R.}~\bibnamefont
  {Aaij}} \emph {et~al.} (\bibinfo {collaboration} {LHCb}),\ }\href {\doibase
  10.1103/PhysRevLett.120.171802} {\bibfield  {journal} {\bibinfo  {journal}
  {Phys. Rev. Lett.}\ }\textbf {\bibinfo {volume} {120}},\ \bibinfo {pages}
  {171802} (\bibinfo {year} {2018}{\natexlab{a}})},\ \Eprint
  {http://arxiv.org/abs/1708.08856} {arXiv:1708.08856 [hep-ex]} \BibitemShut
  {NoStop}%
\bibitem [{\citenamefont {Hirose}\ \emph {et~al.}(2018)\citenamefont {Hirose}
  \emph {et~al.}}]{Belle:2017ilt}%
  \BibitemOpen
  \bibfield  {author} {\bibinfo {author} {\bibfnamefont {S.}~\bibnamefont
  {Hirose}} \emph {et~al.} (\bibinfo {collaboration} {Belle}),\ }\href
  {\doibase 10.1103/PhysRevD.97.012004} {\bibfield  {journal} {\bibinfo
  {journal} {Phys. Rev. D}\ }\textbf {\bibinfo {volume} {97}},\ \bibinfo
  {pages} {012004} (\bibinfo {year} {2018})},\ \Eprint
  {http://arxiv.org/abs/1709.00129} {arXiv:1709.00129 [hep-ex]} \BibitemShut
  {NoStop}%
\bibitem [{\citenamefont {Aaij}\ \emph
  {et~al.}(2018{\natexlab{b}})\citenamefont {Aaij} \emph
  {et~al.}}]{LHCb:2017rln}%
  \BibitemOpen
  \bibfield  {author} {\bibinfo {author} {\bibfnamefont {R.}~\bibnamefont
  {Aaij}} \emph {et~al.} (\bibinfo {collaboration} {LHCb}),\ }\href {\doibase
  10.1103/PhysRevD.97.072013} {\bibfield  {journal} {\bibinfo  {journal} {Phys.
  Rev. D}\ }\textbf {\bibinfo {volume} {97}},\ \bibinfo {pages} {072013}
  (\bibinfo {year} {2018}{\natexlab{b}})},\ \Eprint
  {http://arxiv.org/abs/1711.02505} {arXiv:1711.02505 [hep-ex]} \BibitemShut
  {NoStop}%
\bibitem [{\citenamefont {Abdesselam}\ \emph {et~al.}(2019)\citenamefont
  {Abdesselam} \emph {et~al.}}]{Belle:2019gij}%
  \BibitemOpen
  \bibfield  {author} {\bibinfo {author} {\bibfnamefont {A.}~\bibnamefont
  {Abdesselam}} \emph {et~al.} (\bibinfo {collaboration} {Belle}),\ }\href@noop
  {} {\  (\bibinfo {year} {2019})},\ \Eprint {http://arxiv.org/abs/1904.08794}
  {arXiv:1904.08794 [hep-ex]} \BibitemShut {NoStop}%
\bibitem [{\citenamefont {Amhis}\ \emph {et~al.}(2021)\citenamefont {Amhis}
  \emph {et~al.}}]{HFLAV:2019otj}%
  \BibitemOpen
  \bibfield  {author} {\bibinfo {author} {\bibfnamefont {Y.~S.}\ \bibnamefont
  {Amhis}} \emph {et~al.} (\bibinfo {collaboration} {HFLAV}),\ }\href {\doibase
  10.1140/epjc/s10052-020-8156-7} {\bibfield  {journal} {\bibinfo  {journal}
  {Eur. Phys. J. C}\ }\textbf {\bibinfo {volume} {81}},\ \bibinfo {pages} {226}
  (\bibinfo {year} {2021})},\ \Eprint {http://arxiv.org/abs/1909.12524}
  {arXiv:1909.12524 [hep-ex]} \BibitemShut {NoStop}%
\bibitem [{\citenamefont {Aaij}\ \emph
  {et~al.}(2018{\natexlab{c}})\citenamefont {Aaij} \emph
  {et~al.}}]{LHCb:2017vlu}%
  \BibitemOpen
  \bibfield  {author} {\bibinfo {author} {\bibfnamefont {R.}~\bibnamefont
  {Aaij}} \emph {et~al.} (\bibinfo {collaboration} {LHCb}),\ }\href {\doibase
  10.1103/PhysRevLett.120.121801} {\bibfield  {journal} {\bibinfo  {journal}
  {Phys. Rev. Lett.}\ }\textbf {\bibinfo {volume} {120}},\ \bibinfo {pages}
  {121801} (\bibinfo {year} {2018}{\natexlab{c}})},\ \Eprint
  {http://arxiv.org/abs/1711.05623} {arXiv:1711.05623 [hep-ex]} \BibitemShut
  {NoStop}%
\bibitem [{\citenamefont {Watanabe}(2018)}]{Watanabe:2017mip}%
  \BibitemOpen
  \bibfield  {author} {\bibinfo {author} {\bibfnamefont {R.}~\bibnamefont
  {Watanabe}},\ }\href {\doibase 10.1016/j.physletb.2017.11.016} {\bibfield
  {journal} {\bibinfo  {journal} {Phys. Lett. B}\ }\textbf {\bibinfo {volume}
  {776}},\ \bibinfo {pages} {5} (\bibinfo {year} {2018})},\ \Eprint
  {http://arxiv.org/abs/1709.08644} {arXiv:1709.08644 [hep-ph]} \BibitemShut
  {NoStop}%
\bibitem [{\citenamefont {He}\ and\ \citenamefont
  {Valencia}(2013)}]{He:2012zp}%
  \BibitemOpen
  \bibfield  {author} {\bibinfo {author} {\bibfnamefont {X.-G.}\ \bibnamefont
  {He}}\ and\ \bibinfo {author} {\bibfnamefont {G.}~\bibnamefont {Valencia}},\
  }\href {\doibase 10.1103/PhysRevD.87.014014} {\bibfield  {journal} {\bibinfo
  {journal} {Phys. Rev. D}\ }\textbf {\bibinfo {volume} {87}},\ \bibinfo
  {pages} {014014} (\bibinfo {year} {2013})},\ \Eprint
  {http://arxiv.org/abs/1211.0348} {arXiv:1211.0348 [hep-ph]} \BibitemShut
  {NoStop}%
\bibitem [{\citenamefont {Cveti\v{c}}\ \emph {et~al.}(2017)\citenamefont
  {Cveti\v{c}}, \citenamefont {Halzen}, \citenamefont {Kim},\ and\
  \citenamefont {Oh}}]{Cvetic:2017gkt}%
  \BibitemOpen
  \bibfield  {author} {\bibinfo {author} {\bibfnamefont {G.}~\bibnamefont
  {Cveti\v{c}}}, \bibinfo {author} {\bibfnamefont {F.}~\bibnamefont {Halzen}},
  \bibinfo {author} {\bibfnamefont {C.~S.}\ \bibnamefont {Kim}}, \ and\
  \bibinfo {author} {\bibfnamefont {S.}~\bibnamefont {Oh}},\ }\href {\doibase
  10.1088/1674-1137/41/11/113102} {\bibfield  {journal} {\bibinfo  {journal}
  {Chin. Phys. C}\ }\textbf {\bibinfo {volume} {41}},\ \bibinfo {pages}
  {113102} (\bibinfo {year} {2017})},\ \Eprint
  {http://arxiv.org/abs/1702.04335} {arXiv:1702.04335 [hep-ph]} \BibitemShut
  {NoStop}%
\bibitem [{\citenamefont {Asadi}\ \emph {et~al.}(2018)\citenamefont {Asadi},
  \citenamefont {Buckley},\ and\ \citenamefont {Shih}}]{Asadi:2018wea}%
  \BibitemOpen
  \bibfield  {author} {\bibinfo {author} {\bibfnamefont {P.}~\bibnamefont
  {Asadi}}, \bibinfo {author} {\bibfnamefont {M.~R.}\ \bibnamefont {Buckley}},
  \ and\ \bibinfo {author} {\bibfnamefont {D.}~\bibnamefont {Shih}},\ }\href
  {\doibase 10.1007/JHEP09(2018)010} {\bibfield  {journal} {\bibinfo  {journal}
  {JHEP}\ }\textbf {\bibinfo {volume} {09}},\ \bibinfo {pages} {010} (\bibinfo
  {year} {2018})},\ \Eprint {http://arxiv.org/abs/1804.04135} {arXiv:1804.04135
  [hep-ph]} \BibitemShut {NoStop}%
\bibitem [{\citenamefont {Greljo}\ \emph {et~al.}(2018)\citenamefont {Greljo},
  \citenamefont {Robinson}, \citenamefont {Shakya},\ and\ \citenamefont
  {Zupan}}]{Greljo:2018ogz}%
  \BibitemOpen
  \bibfield  {author} {\bibinfo {author} {\bibfnamefont {A.}~\bibnamefont
  {Greljo}}, \bibinfo {author} {\bibfnamefont {D.~J.}\ \bibnamefont
  {Robinson}}, \bibinfo {author} {\bibfnamefont {B.}~\bibnamefont {Shakya}}, \
  and\ \bibinfo {author} {\bibfnamefont {J.}~\bibnamefont {Zupan}},\ }\href
  {\doibase 10.1007/JHEP09(2018)169} {\bibfield  {journal} {\bibinfo  {journal}
  {JHEP}\ }\textbf {\bibinfo {volume} {09}},\ \bibinfo {pages} {169} (\bibinfo
  {year} {2018})},\ \Eprint {http://arxiv.org/abs/1804.04642} {arXiv:1804.04642
  [hep-ph]} \BibitemShut {NoStop}%
\bibitem [{\citenamefont {Babu}\ \emph {et~al.}(2019)\citenamefont {Babu},
  \citenamefont {Dutta},\ and\ \citenamefont {Mohapatra}}]{Babu:2018vrl}%
  \BibitemOpen
  \bibfield  {author} {\bibinfo {author} {\bibfnamefont {K.~S.}\ \bibnamefont
  {Babu}}, \bibinfo {author} {\bibfnamefont {B.}~\bibnamefont {Dutta}}, \ and\
  \bibinfo {author} {\bibfnamefont {R.~N.}\ \bibnamefont {Mohapatra}},\ }\href
  {\doibase 10.1007/JHEP01(2019)168} {\bibfield  {journal} {\bibinfo  {journal}
  {JHEP}\ }\textbf {\bibinfo {volume} {01}},\ \bibinfo {pages} {168} (\bibinfo
  {year} {2019})},\ \Eprint {http://arxiv.org/abs/1811.04496} {arXiv:1811.04496
  [hep-ph]} \BibitemShut {NoStop}%
\bibitem [{\citenamefont {Mandal}\ \emph {et~al.}(2020)\citenamefont {Mandal},
  \citenamefont {Murgui}, \citenamefont {Pe\~nuelas},\ and\ \citenamefont
  {Pich}}]{Mandal:2020htr}%
  \BibitemOpen
  \bibfield  {author} {\bibinfo {author} {\bibfnamefont {R.}~\bibnamefont
  {Mandal}}, \bibinfo {author} {\bibfnamefont {C.}~\bibnamefont {Murgui}},
  \bibinfo {author} {\bibfnamefont {A.}~\bibnamefont {Pe\~nuelas}}, \ and\
  \bibinfo {author} {\bibfnamefont {A.}~\bibnamefont {Pich}},\ }\href {\doibase
  10.1007/JHEP08(2020)022} {\bibfield  {journal} {\bibinfo  {journal} {JHEP}\
  }\textbf {\bibinfo {volume} {08}},\ \bibinfo {pages} {022} (\bibinfo {year}
  {2020})},\ \Eprint {http://arxiv.org/abs/2004.06726} {arXiv:2004.06726
  [hep-ph]} \BibitemShut {NoStop}%
\bibitem [{\citenamefont {Abi}\ \emph {et~al.}(2021)\citenamefont {Abi} \emph
  {et~al.}}]{Muong-2:2021ojo}%
  \BibitemOpen
  \bibfield  {author} {\bibinfo {author} {\bibfnamefont {B.}~\bibnamefont
  {Abi}} \emph {et~al.} (\bibinfo {collaboration} {Muon g-2}),\ }\href
  {\doibase 10.1103/PhysRevLett.126.141801} {\bibfield  {journal} {\bibinfo
  {journal} {Phys. Rev. Lett.}\ }\textbf {\bibinfo {volume} {126}},\ \bibinfo
  {pages} {141801} (\bibinfo {year} {2021})},\ \Eprint
  {http://arxiv.org/abs/2104.03281} {arXiv:2104.03281 [hep-ex]} \BibitemShut
  {NoStop}%
\bibitem [{\citenamefont {Aaij}\ \emph {et~al.}(2021)\citenamefont {Aaij} \emph
  {et~al.}}]{LHCb:2021trn}%
  \BibitemOpen
  \bibfield  {author} {\bibinfo {author} {\bibfnamefont {R.}~\bibnamefont
  {Aaij}} \emph {et~al.} (\bibinfo {collaboration} {LHCb}),\ }\href@noop {} {\
  (\bibinfo {year} {2021})},\ \Eprint {http://arxiv.org/abs/2103.11769}
  {arXiv:2103.11769 [hep-ex]} \BibitemShut {NoStop}%
\bibitem [{\citenamefont {Duraisamy}\ and\ \citenamefont
  {Datta}(2013)}]{Duraisamy:2013pia}%
  \BibitemOpen
  \bibfield  {author} {\bibinfo {author} {\bibfnamefont {M.}~\bibnamefont
  {Duraisamy}}\ and\ \bibinfo {author} {\bibfnamefont {A.}~\bibnamefont
  {Datta}},\ }\href {\doibase 10.1007/JHEP09(2013)059} {\bibfield  {journal}
  {\bibinfo  {journal} {JHEP}\ }\textbf {\bibinfo {volume} {09}},\ \bibinfo
  {pages} {059} (\bibinfo {year} {2013})},\ \Eprint
  {http://arxiv.org/abs/1302.7031} {arXiv:1302.7031 [hep-ph]} \BibitemShut
  {NoStop}%
\bibitem [{\citenamefont {Duraisamy}\ \emph {et~al.}(2014)\citenamefont
  {Duraisamy}, \citenamefont {Sharma},\ and\ \citenamefont
  {Datta}}]{Duraisamy:2014sna}%
  \BibitemOpen
  \bibfield  {author} {\bibinfo {author} {\bibfnamefont {M.}~\bibnamefont
  {Duraisamy}}, \bibinfo {author} {\bibfnamefont {P.}~\bibnamefont {Sharma}}, \
  and\ \bibinfo {author} {\bibfnamefont {A.}~\bibnamefont {Datta}},\ }\href
  {\doibase 10.1103/PhysRevD.90.074013} {\bibfield  {journal} {\bibinfo
  {journal} {Phys. Rev. D}\ }\textbf {\bibinfo {volume} {90}},\ \bibinfo
  {pages} {074013} (\bibinfo {year} {2014})},\ \Eprint
  {http://arxiv.org/abs/1405.3719} {arXiv:1405.3719 [hep-ph]} \BibitemShut
  {NoStop}%
\bibitem [{\citenamefont {Bhattacharya}\ \emph {et~al.}(2022)\citenamefont
  {Bhattacharya}, \citenamefont {Browder}, \citenamefont {Campagna},
  \citenamefont {Datta}, \citenamefont {Dubey}, \citenamefont {Mukherjee},\
  and\ \citenamefont {Sibidanov}}]{Bhattacharya:2022cna}%
  \BibitemOpen
  \bibfield  {author} {\bibinfo {author} {\bibfnamefont {B.}~\bibnamefont
  {Bhattacharya}}, \bibinfo {author} {\bibfnamefont {T.}~\bibnamefont
  {Browder}}, \bibinfo {author} {\bibfnamefont {Q.}~\bibnamefont {Campagna}},
  \bibinfo {author} {\bibfnamefont {A.}~\bibnamefont {Datta}}, \bibinfo
  {author} {\bibfnamefont {S.}~\bibnamefont {Dubey}}, \bibinfo {author}
  {\bibfnamefont {L.}~\bibnamefont {Mukherjee}}, \ and\ \bibinfo {author}
  {\bibfnamefont {A.}~\bibnamefont {Sibidanov}},\ }in\ \href@noop {} {\emph
  {\bibinfo {booktitle} {{2022 Snowmass Summer Study}}}}\ (\bibinfo {year}
  {2022})\ \Eprint {http://arxiv.org/abs/2203.07189} {arXiv:2203.07189
  [hep-ph]} \BibitemShut {NoStop}%
\bibitem [{\citenamefont {Waheed}\ \emph {et~al.}(2019)\citenamefont {Waheed}
  \emph {et~al.}}]{Belle:2018ezy}%
  \BibitemOpen
  \bibfield  {author} {\bibinfo {author} {\bibfnamefont {E.}~\bibnamefont
  {Waheed}} \emph {et~al.} (\bibinfo {collaboration} {Belle}),\ }\href
  {\doibase 10.1103/PhysRevD.100.052007} {\bibfield  {journal} {\bibinfo
  {journal} {Phys. Rev. D}\ }\textbf {\bibinfo {volume} {100}},\ \bibinfo
  {pages} {052007} (\bibinfo {year} {2019})},\ \bibinfo {note} {[Erratum:
  Phys.Rev.D 103, 079901 (2021)]},\ \Eprint {http://arxiv.org/abs/1809.03290}
  {arXiv:1809.03290 [hep-ex]} \BibitemShut {NoStop}%
\bibitem [{\citenamefont {Bobeth}\ \emph {et~al.}(2021)\citenamefont {Bobeth},
  \citenamefont {van Dyk}, \citenamefont {Bordone}, \citenamefont {Jung},\ and\
  \citenamefont {Gubernari}}]{Bobeth:2021lya}%
  \BibitemOpen
  \bibfield  {author} {\bibinfo {author} {\bibfnamefont {C.}~\bibnamefont
  {Bobeth}}, \bibinfo {author} {\bibfnamefont {D.}~\bibnamefont {van Dyk}},
  \bibinfo {author} {\bibfnamefont {M.}~\bibnamefont {Bordone}}, \bibinfo
  {author} {\bibfnamefont {M.}~\bibnamefont {Jung}}, \ and\ \bibinfo {author}
  {\bibfnamefont {N.}~\bibnamefont {Gubernari}},\ }\href@noop {} {\  (\bibinfo
  {year} {2021})},\ \Eprint {http://arxiv.org/abs/2104.02094} {arXiv:2104.02094
  [hep-ph]} \BibitemShut {NoStop}%
\bibitem [{\citenamefont {Carvunis}\ \emph {et~al.}(2022)\citenamefont
  {Carvunis}, \citenamefont {Crivellin}, \citenamefont {Guadagnoli},\ and\
  \citenamefont {Gangal}}]{Carvunis:2021dss}%
  \BibitemOpen
  \bibfield  {author} {\bibinfo {author} {\bibfnamefont {A.}~\bibnamefont
  {Carvunis}}, \bibinfo {author} {\bibfnamefont {A.}~\bibnamefont {Crivellin}},
  \bibinfo {author} {\bibfnamefont {D.}~\bibnamefont {Guadagnoli}}, \ and\
  \bibinfo {author} {\bibfnamefont {S.}~\bibnamefont {Gangal}},\ }\href
  {\doibase 10.1103/PhysRevD.105.L031701} {\bibfield  {journal} {\bibinfo
  {journal} {Phys. Rev. D}\ }\textbf {\bibinfo {volume} {105}},\ \bibinfo
  {pages} {L031701} (\bibinfo {year} {2022})},\ \Eprint
  {http://arxiv.org/abs/2106.09610} {arXiv:2106.09610 [hep-ph]} \BibitemShut
  {NoStop}%
\bibitem [{\citenamefont {Grzadkowski}\ \emph {et~al.}(2010)\citenamefont
  {Grzadkowski}, \citenamefont {Iskrzynski}, \citenamefont {Misiak},\ and\
  \citenamefont {Rosiek}}]{Grzadkowski:2010es}%
  \BibitemOpen
  \bibfield  {author} {\bibinfo {author} {\bibfnamefont {B.}~\bibnamefont
  {Grzadkowski}}, \bibinfo {author} {\bibfnamefont {M.}~\bibnamefont
  {Iskrzynski}}, \bibinfo {author} {\bibfnamefont {M.}~\bibnamefont {Misiak}},
  \ and\ \bibinfo {author} {\bibfnamefont {J.}~\bibnamefont {Rosiek}},\ }\href
  {\doibase 10.1007/JHEP10(2010)085} {\bibfield  {journal} {\bibinfo  {journal}
  {JHEP}\ }\textbf {\bibinfo {volume} {10}},\ \bibinfo {pages} {085} (\bibinfo
  {year} {2010})},\ \Eprint {http://arxiv.org/abs/1008.4884} {arXiv:1008.4884
  [hep-ph]} \BibitemShut {NoStop}%
\bibitem [{\citenamefont {Henning}\ \emph {et~al.}(2016)\citenamefont
  {Henning}, \citenamefont {Lu},\ and\ \citenamefont
  {Murayama}}]{Henning:2014wua}%
  \BibitemOpen
  \bibfield  {author} {\bibinfo {author} {\bibfnamefont {B.}~\bibnamefont
  {Henning}}, \bibinfo {author} {\bibfnamefont {X.}~\bibnamefont {Lu}}, \ and\
  \bibinfo {author} {\bibfnamefont {H.}~\bibnamefont {Murayama}},\ }\href
  {\doibase 10.1007/JHEP01(2016)023} {\bibfield  {journal} {\bibinfo  {journal}
  {JHEP}\ }\textbf {\bibinfo {volume} {01}},\ \bibinfo {pages} {023} (\bibinfo
  {year} {2016})},\ \Eprint {http://arxiv.org/abs/1412.1837} {arXiv:1412.1837
  [hep-ph]} \BibitemShut {NoStop}%
\bibitem [{\citenamefont {Brivio}\ and\ \citenamefont
  {Trott}(2019)}]{Brivio:2017vri}%
  \BibitemOpen
  \bibfield  {author} {\bibinfo {author} {\bibfnamefont {I.}~\bibnamefont
  {Brivio}}\ and\ \bibinfo {author} {\bibfnamefont {M.}~\bibnamefont {Trott}},\
  }\href {\doibase 10.1016/j.physrep.2018.11.002} {\bibfield  {journal}
  {\bibinfo  {journal} {Phys. Rept.}\ }\textbf {\bibinfo {volume} {793}},\
  \bibinfo {pages} {1} (\bibinfo {year} {2019})},\ \Eprint
  {http://arxiv.org/abs/1706.08945} {arXiv:1706.08945 [hep-ph]} \BibitemShut
  {NoStop}%
\bibitem [{\citenamefont {del Aguila}\ \emph {et~al.}(2009)\citenamefont {del
  Aguila}, \citenamefont {Bar-Shalom}, \citenamefont {Soni},\ and\
  \citenamefont {Wudka}}]{delAguila:2008ir}%
  \BibitemOpen
  \bibfield  {author} {\bibinfo {author} {\bibfnamefont {F.}~\bibnamefont {del
  Aguila}}, \bibinfo {author} {\bibfnamefont {S.}~\bibnamefont {Bar-Shalom}},
  \bibinfo {author} {\bibfnamefont {A.}~\bibnamefont {Soni}}, \ and\ \bibinfo
  {author} {\bibfnamefont {J.}~\bibnamefont {Wudka}},\ }\href {\doibase
  10.1016/j.physletb.2008.11.031} {\bibfield  {journal} {\bibinfo  {journal}
  {Phys. Lett. B}\ }\textbf {\bibinfo {volume} {670}},\ \bibinfo {pages} {399}
  (\bibinfo {year} {2009})},\ \Eprint {http://arxiv.org/abs/0806.0876}
  {arXiv:0806.0876 [hep-ph]} \BibitemShut {NoStop}%
\bibitem [{\citenamefont {Aparici}\ \emph {et~al.}(2009)\citenamefont
  {Aparici}, \citenamefont {Kim}, \citenamefont {Santamaria},\ and\
  \citenamefont {Wudka}}]{Aparici:2009fh}%
  \BibitemOpen
  \bibfield  {author} {\bibinfo {author} {\bibfnamefont {A.}~\bibnamefont
  {Aparici}}, \bibinfo {author} {\bibfnamefont {K.}~\bibnamefont {Kim}},
  \bibinfo {author} {\bibfnamefont {A.}~\bibnamefont {Santamaria}}, \ and\
  \bibinfo {author} {\bibfnamefont {J.}~\bibnamefont {Wudka}},\ }\href
  {\doibase 10.1103/PhysRevD.80.013010} {\bibfield  {journal} {\bibinfo
  {journal} {Phys. Rev. D}\ }\textbf {\bibinfo {volume} {80}},\ \bibinfo
  {pages} {013010} (\bibinfo {year} {2009})},\ \Eprint
  {http://arxiv.org/abs/0904.3244} {arXiv:0904.3244 [hep-ph]} \BibitemShut
  {NoStop}%
\bibitem [{\citenamefont {Bhattacharya}\ and\ \citenamefont
  {Wudka}(2016)}]{Bhattacharya:2015vja}%
  \BibitemOpen
  \bibfield  {author} {\bibinfo {author} {\bibfnamefont {S.}~\bibnamefont
  {Bhattacharya}}\ and\ \bibinfo {author} {\bibfnamefont {J.}~\bibnamefont
  {Wudka}},\ }\href {\doibase 10.1103/PhysRevD.94.055022} {\bibfield  {journal}
  {\bibinfo  {journal} {Phys. Rev. D}\ }\textbf {\bibinfo {volume} {94}},\
  \bibinfo {pages} {055022} (\bibinfo {year} {2016})},\ \bibinfo {note}
  {[Erratum: Phys.Rev.D 95, 039904 (2017)]},\ \Eprint
  {http://arxiv.org/abs/1505.05264} {arXiv:1505.05264 [hep-ph]} \BibitemShut
  {NoStop}%
\bibitem [{\citenamefont {Liao}\ and\ \citenamefont {Ma}(2017)}]{Liao:2016qyd}%
  \BibitemOpen
  \bibfield  {author} {\bibinfo {author} {\bibfnamefont {Y.}~\bibnamefont
  {Liao}}\ and\ \bibinfo {author} {\bibfnamefont {X.-D.}\ \bibnamefont {Ma}},\
  }\href {\doibase 10.1103/PhysRevD.96.015012} {\bibfield  {journal} {\bibinfo
  {journal} {Phys. Rev. D}\ }\textbf {\bibinfo {volume} {96}},\ \bibinfo
  {pages} {015012} (\bibinfo {year} {2017})},\ \Eprint
  {http://arxiv.org/abs/1612.04527} {arXiv:1612.04527 [hep-ph]} \BibitemShut
  {NoStop}%
\bibitem [{\citenamefont {Bischer}\ and\ \citenamefont
  {Rodejohann}(2019)}]{Bischer:2019ttk}%
  \BibitemOpen
  \bibfield  {author} {\bibinfo {author} {\bibfnamefont {I.}~\bibnamefont
  {Bischer}}\ and\ \bibinfo {author} {\bibfnamefont {W.}~\bibnamefont
  {Rodejohann}},\ }\href {\doibase 10.1016/j.nuclphysb.2019.114746} {\bibfield
  {journal} {\bibinfo  {journal} {Nucl. Phys. B}\ }\textbf {\bibinfo {volume}
  {947}},\ \bibinfo {pages} {114746} (\bibinfo {year} {2019})},\ \Eprint
  {http://arxiv.org/abs/1905.08699} {arXiv:1905.08699 [hep-ph]} \BibitemShut
  {NoStop}%
\bibitem [{\citenamefont {Datta}\ \emph
  {et~al.}(2021{\natexlab{a}})\citenamefont {Datta}, \citenamefont {Kumar},
  \citenamefont {Liu},\ and\ \citenamefont {Marfatia}}]{Datta:2020ocb}%
  \BibitemOpen
  \bibfield  {author} {\bibinfo {author} {\bibfnamefont {A.}~\bibnamefont
  {Datta}}, \bibinfo {author} {\bibfnamefont {J.}~\bibnamefont {Kumar}},
  \bibinfo {author} {\bibfnamefont {H.}~\bibnamefont {Liu}}, \ and\ \bibinfo
  {author} {\bibfnamefont {D.}~\bibnamefont {Marfatia}},\ }\href {\doibase
  10.1007/JHEP02(2021)015} {\bibfield  {journal} {\bibinfo  {journal} {JHEP}\
  }\textbf {\bibinfo {volume} {02}},\ \bibinfo {pages} {015} (\bibinfo {year}
  {2021}{\natexlab{a}})},\ \Eprint {http://arxiv.org/abs/2010.12109}
  {arXiv:2010.12109 [hep-ph]} \BibitemShut {NoStop}%
\bibitem [{\citenamefont {Datta}\ \emph
  {et~al.}(2021{\natexlab{b}})\citenamefont {Datta}, \citenamefont {Kumar},
  \citenamefont {Liu},\ and\ \citenamefont {Marfatia}}]{Datta:2021akg}%
  \BibitemOpen
  \bibfield  {author} {\bibinfo {author} {\bibfnamefont {A.}~\bibnamefont
  {Datta}}, \bibinfo {author} {\bibfnamefont {J.}~\bibnamefont {Kumar}},
  \bibinfo {author} {\bibfnamefont {H.}~\bibnamefont {Liu}}, \ and\ \bibinfo
  {author} {\bibfnamefont {D.}~\bibnamefont {Marfatia}},\ }\href {\doibase
  10.1007/JHEP05(2021)037} {\bibfield  {journal} {\bibinfo  {journal} {JHEP}\
  }\textbf {\bibinfo {volume} {05}},\ \bibinfo {pages} {037} (\bibinfo {year}
  {2021}{\natexlab{b}})},\ \Eprint {http://arxiv.org/abs/2103.04441}
  {arXiv:2103.04441 [hep-ph]} \BibitemShut {NoStop}%
\bibitem [{\citenamefont {Bordone}\ \emph {et~al.}(2020)\citenamefont
  {Bordone}, \citenamefont {Gubernari}, \citenamefont {van Dyk},\ and\
  \citenamefont {Jung}}]{Bordone:2019guc}%
  \BibitemOpen
  \bibfield  {author} {\bibinfo {author} {\bibfnamefont {M.}~\bibnamefont
  {Bordone}}, \bibinfo {author} {\bibfnamefont {N.}~\bibnamefont {Gubernari}},
  \bibinfo {author} {\bibfnamefont {D.}~\bibnamefont {van Dyk}}, \ and\
  \bibinfo {author} {\bibfnamefont {M.}~\bibnamefont {Jung}},\ }\href {\doibase
  10.1140/epjc/s10052-020-7850-9} {\bibfield  {journal} {\bibinfo  {journal}
  {Eur. Phys. J. C}\ }\textbf {\bibinfo {volume} {80}},\ \bibinfo {pages} {347}
  (\bibinfo {year} {2020})},\ \Eprint {http://arxiv.org/abs/1912.09335}
  {arXiv:1912.09335 [hep-ph]} \BibitemShut {NoStop}%
\bibitem [{\citenamefont {Tanaka}\ and\ \citenamefont
  {Watanabe}(2013)}]{Tanaka:2012nw}%
  \BibitemOpen
  \bibfield  {author} {\bibinfo {author} {\bibfnamefont {M.}~\bibnamefont
  {Tanaka}}\ and\ \bibinfo {author} {\bibfnamefont {R.}~\bibnamefont
  {Watanabe}},\ }\href {\doibase 10.1103/PhysRevD.87.034028} {\bibfield
  {journal} {\bibinfo  {journal} {Phys. Rev. D}\ }\textbf {\bibinfo {volume}
  {87}},\ \bibinfo {pages} {034028} (\bibinfo {year} {2013})},\ \Eprint
  {http://arxiv.org/abs/1212.1878} {arXiv:1212.1878 [hep-ph]} \BibitemShut
  {NoStop}%
\bibitem [{\citenamefont {Iguro}\ and\ \citenamefont
  {Watanabe}(2020)}]{Iguro:2020cpg}%
  \BibitemOpen
  \bibfield  {author} {\bibinfo {author} {\bibfnamefont {S.}~\bibnamefont
  {Iguro}}\ and\ \bibinfo {author} {\bibfnamefont {R.}~\bibnamefont
  {Watanabe}},\ }\href {\doibase 10.1007/JHEP08(2020)006} {\bibfield  {journal}
  {\bibinfo  {journal} {JHEP}\ }\textbf {\bibinfo {volume} {08}},\ \bibinfo
  {pages} {006} (\bibinfo {year} {2020})},\ \Eprint
  {http://arxiv.org/abs/2004.10208} {arXiv:2004.10208 [hep-ph]} \BibitemShut
  {NoStop}%
\end{thebibliography}%
